\documentclass[12pt]{article}
\usepackage{feynmp}
\usepackage{amssymb,graphicx}
\usepackage{epstopdf}
\usepackage{tabularx}
\usepackage{array}
\usepackage{subcaption}
\usepackage{amsmath,amsfonts}
\usepackage{epsfig}
\usepackage{xcolor}
\usepackage{float}
\usepackage{cite}
\usepackage[normalem]{ulem}

\newcommand{\be}{\begin{equation}}
\newcommand{\ee}{\end{equation}}
\newcommand{\bea}{\begin{eqnarray}}
\newcommand{\eea}{\end{eqnarray}}
\newcommand{\bg}{\begin{gather}}
\newcommand{\eg}{\end{gather}}
\newcommand{\bseq}{\begin{subequations}}
\newcommand{\eseq}{\end{subequations}}

\renewcommand{\ln}{\mathop{\rm ln}\nolimits}

\begin{document}
    
\begin{flushright}
INR-TH-2022-014
\end{flushright}
\vspace{10pt}
\begin{center}
  {\LARGE \bf   Generating cosmological perturbations \\in non-singular Horndeski cosmologies} \\
\vspace{20pt}
Y.~Ageeva$^{a,b,c,}$\footnote[1]{{\bf email:}
    ageeva@inr.ac.ru}, P. Petrov$^{a,}$\footnote[2]{{\bf email:}
    petrov@inr.ac.ru},
  V. Rubakov$^{a,b,}$\footnote[3]{{\bf email:} rubakov@inr.ac.ru}\\
\vspace{15pt}
  $^a$\textit{
Institute for Nuclear Research of
         the Russian Academy of Sciences,\\  60th October Anniversary
  Prospect, 7a, 117312 Moscow, Russia}\\
\vspace{5pt}
$^b$\textit{Department of Particle Physics and Cosmology,\\
  Physics Faculty, M.V.~Lomonosov
  Moscow State University, \\Leninskie Gory 1-2,  119991 Moscow, Russia
  }\\
\vspace{5pt}
$^c$\textit{
  Institute for Theoretical and Mathematical Physics,\\
  M.V.~Lomonosov Moscow State University,\\ Leninskie Gory 1,
119991 Moscow,
Russia
}
    \end{center}
    \vspace{5pt}
\begin{abstract}
We construct a concrete model of Horndeski bounce with 
 strong gravity in the past. Within this model we show that the 
 correct spectra of cosmological perturbations may be generated at
 early contracting epoch, with mild
fine-tuning ensuring that the scalar spectral 
tilt
$n_S$
and tensor-to-scalar ratio $r$ are consistent with observations.
The smallness
of $r$ is governed by the smallness of the scalar sound 
speed. Arbitrarily small values of $r$ are forbidden in our setup because of
the
strong coupling in the past. Nevertheless,  
we show that
it is possible to generate perturbations in a controllable 
way, i.e. in the regime where the background evolution 
and perturbations are legitimately described within classical 
field theory and weakly coupled quantum theory.

\end{abstract}    
    
\section{Introduction}
\label{sec:intro}

Nowadays a vigorous research of non-singular scenarios which are alternative or complementary to 
inflation continues. The realization of non-singular behaviour within classical field theory requires
the violation of the null convergence condition (null energy condition (NEC) in General
Relativity), see Ref.~\cite{Rubakov:2014jja} for a review. However, models with unusual matter which violates the NEC or null convergence condition \cite{Tipler:1978zz} often suffer from pathological behavior because of various kinds of instabilities.
In Refs.~\cite{Creminelli:2010ba,Deffayet:2010qz,Kobayashi:2010cm} it was first shown,
that stable violation of NEC can be implemented in Horndeski
theories~\cite{Horndeski:1974wa} (for  reviews see, e.g.,
Refs. \cite{Rubakov:2014jja,Kobayashi:2019hrl}).
Horndeski theory is a scalar-tensor modification of gravity. The Lagrangian of the latter
contains second derivatives of both the metric and
scalar field, but yet provides the
second-order equations of motion. The explicit examples of healthy NEC-violating non-singular early stages, such as bouncing solutions \cite{Novello:2008ra,Lehners:2008vx,Lehners:2011kr,Battefeld:2014uga,Qiu:2011cy,Easson:2011zy,Cai:2012va,Osipov:2013ssa,Qiu:2013eoa,Koehn:2013upa,Battarra:2014tga,Qiu:2015nha,Ijjas:2016tpn,Brandenberger:2016vhg,Ijjas:2018qbo,Kobayashi:2019hrl,Mironov:2019haz}
or Galilean genesis \cite{Creminelli:2010ba,Creminelli:2012my,Hinterbichler:2012fr,Elder:2013gya,Pirtskhalava:2014esa,Nishi:2015pta,Kobayashi:2015gga} are numerous in Horndeski 
theory and its subclasses.
In this paper we focus on bouncing Universe.

Yet another problem arises on the way to constructing the whole  cosmological evolution of the Universe. Cosmologies which involve early non-singular epoch typically
suffer from either  singularities or
gradient and/or ghost instabilities at some moment of time in the past or in the future
 (possibly
well before or well after the bounce). Thus, even within Horndeski
theories, extending bouncing cosmology to the whole time
interval
$-\infty<t<+\infty$ is not straightforward. Namely, perturbations about
bouncing spatially flat FLRW
backgrounds $ds^2 = -dt^2 + a^2(t)\delta_{ij} dx^i dx^j$ have either  
gradient or ghost instabilities, or both, provided that the following two integrals diverge:
\begin{subequations}
  \begin{align}
\label{jan21-22-1}
    \int_{-\infty}^{t} a(t) (\mathcal{ F}_T +\mathcal{ F}_S) dt &= \infty \; ,
    \\
   \int_t^{+\infty} a(t) (\mathcal{ F}_T +\mathcal{ F}_S) dt &=
  \infty \; ,
\end{align}
  \end{subequations}
where 
$\mathcal{ F}_T$ and  $\mathcal{ F}_S$ are time-dependent coefficients in the
quadratic
actions for tensor (transverse traceless $h_{ij}$) and scalar ($\zeta$)
perturbations,
\begin{subequations}
  \label{jan21-22-2}
  \begin{align}
      \label{jan21-22-2a}
 \mathcal{ S}_{hh} & =\int dt d^3x \frac{ a^3}{8}\left[
        \mathcal{ G}_T
        \dot h_{ij}^2
        -\frac{\mathcal{ F}_T}{a^2}
        h_{ij,k} h_{ij,k} \right] \; ,
 \\
   \mathcal{ S}_{\zeta\zeta} &=\int dt d^3x a^3\left[
        \mathcal{ G}_S
        \dot\zeta^2
        -\frac{\mathcal{ F}_S}{a^2}
        \zeta_{,i}\zeta_{,i}
        \right] \; .
\end{align}
\end{subequations}
This property, known as the no-go
theorem~\cite{Libanov:2016kfc,Kobayashi:2016xpl,Kolevatov:2016ppi,Akama:2017jsa},
rules out the simplest bouncing Horndeski setups where $a(t)$ tends to infinity
while  $\mathcal{ F}_T$ and  $\mathcal{ F}_S$ stay positive and
finite
as $t \to \pm \infty$.

One
way~\cite{Cai:2017dyi,Kolevatov:2017voe,Ye:2019sth,Mironov:2019qjt,Ilyas:2020qja,Zhu:2021whu}
of getting around this
theorem
is to employ beyond Horndeski theories~\cite{Zumalacarregui:2013pma,Gleyzes:2014dya}
Degenerate Higher-Order Scalar-Tensor  (DHOST) generalizations~\cite{Langlois:2015cwa}. These, however, have their
own problems, since adding conventional matter fields often results in
superluminality~\cite{Mironov:2020pqh}. Without employing these generalizations, i.e., 
staying within the Horndeski
class, there is still a possibility  to allow the coefficients
$\mathcal{ G}_T$, $\mathcal{ F}_T$,  $\mathcal{ G}_S$ and  $\mathcal{ F}_S$ to
tend to zero as $t\to -\infty$ in such a way that the integral in the left hand side of
\eqref{jan21-22-1} converges in the lower limit~\cite{Kobayashi:2016xpl}. At first sight
this is dangerous from the viewpoint of strong coupling: the coefficients
$\mathcal{ G}_T$, $\mathcal{ F}_T$,  $\mathcal{ G}_S$ and  $\mathcal{ F}_S$ are
analogs of the Planck mass squared, so their early-time behavior
$\mathcal{ G}_T, \mathcal{ F}_T,  \mathcal{ G}_S, \mathcal{ F}_S \to 0$ as
$t \to -\infty$
implies
that the gravitational interaction is strong in remote past; note that we always work in the
Jordan frame and that $a(t)$ grows backwards in time at early times. Nevertheless,
depending on the model, the cosmological evolution may, in fact, be legitimately described
within classical field theory at all times, since even at early times
the classical energy scale
(which is $|t|^{-1}$ for power-law bounce) may be lower than the quantum strong coupling
scale~\cite{Ageeva:2018lko,Ageeva:2020gti,Ageeva:2020buc,Ageeva:2021yik}.
It is worth emphasizing that this idea of healthy bounce with
``strong gravity in the past''
(meaning that $\mathcal{ G}_T, \mathcal{ F}_T,  \mathcal{ G}_S, \mathcal{ F}_S \to 0$ as
$t \to -\infty$) has been re-invented, albeit not quite explicitly,
in Refs.~\cite{Nandi:2020sif,Nandi:2020szp} from
a different
prospective: bounce in the Jordan frame is obtained 
there
via conformal
transformation from an inflationary setup in the Einstein frame.
Unlike in 
Refs.~\cite{Nandi:2020sif,Nandi:2020szp}, we
work directly in the Jordan frame.
%

It is relatively straightforward to construct Horndeski models which admit bouncing
solutions with power-law asymptotics at early times~\cite{Kobayashi:2016xpl,Ageeva:2021yik},
\be
a(t) \propto (-t)^\chi \;, \;\;\;\;\;
\mathcal{ G}_T, \mathcal{ F}_T,  \mathcal{ G}_S, \mathcal{ F}_S \propto \frac{1}{(-t)^{2\mu}}
\;, \;\;\; t \to -\infty \; ,
\label{jan24-22-1}
\ee
with time-independent parameters $1> \chi> 0$,
$2\mu > \chi +1$.
The latter property guarantees that the inequality \eqref{jan21-22-1}
does not hold, which is a pre-requisite for healthy bounce.
In this paper we concentrate on this
simple case and consider the generation of cosmological perturbations at early
contraction stage.
In terms of conformal time $\eta  \propto - (-t)^{1-\chi}$, the quadratic action
\eqref{jan21-22-2a} for tensor perturbations coincides with that
in General Relativity
with the background scale factor
\be
a_{E} (\eta) = a(\eta)  \mathcal{ G}_T^{1/2}(\eta)
\propto \frac{1}{(-\eta)^{\frac{\mu - \chi}{1-\chi}}} \; .
\label{jan24-22-11}
\ee
In fact, this is precisely
  the scale factor in the Einstein frame in our class of models.
  Now, for $\mu <1$ the
  Einstein frame cosmic time $t_{E} =\int~a_{E}(\eta) d\eta
= - (-\eta)^{\frac{1-\mu}{1-\chi}}$ runs
  from $t_{E} = - \infty$, and  the effective scale factor increases as
  $a_{E} = (-t_{E})^{-b}$ where $b = \frac{\mu - \chi}{1-\mu} > 1$
  ~\cite{Ageeva:2020gti}. Such a geometry is singular as
  $t_{E} \to -\infty$: it is past geodesically incomplete
  and cannot be completed\footnote{This property is not pathological in
    our case,
    since by assumption particles with time-independent mass feel
    the Jordan frame geometry rather than the Einstein frame one.}.
  On the other hand, for
  $\mu >1$ one immediately recognizes  effective power-law inflation with
  \be
  a_{E}(t_E
  ) \propto t_{E}^{\frac{\mu - \chi}{\mu -1}} \; ,
\label{jun23-22-1}
  \ee
  where
  $t_{E} \propto  (-\eta)^{- \frac{\mu -1}{1-\chi}}$ runs from $t_{E} =0$.
  In the Einstein frame, this is a version of
  $G$-inflation~\cite{Kobayashi:2010cm}. In either case,
  for $\mu \approx 1$ the Einstein frame expansion is nearly exponential, so
  the power spectrum of generated tensor perturbations
  is nearly flat;
  similar observation applies to scalar perturbations as well.
  This implies that Horndeski bounce with ``strong gravity in the past'' may be capable
  of generating realistic cosmological perturbations; we again emphasize the
  similarity with Refs.~\cite{Nandi:2020sif,Nandi:2020szp,Kobayashi:2010cm}
  (where approximate flatness
  of the spectra is built in by construction).

  In this paper we consider the models from the class of
  Refs.~\cite{Kobayashi:2016xpl,Ageeva:2021yik}, as described
  below.
    The issues we would like to understand are: (i) what governs
  spectral tilts and the overall amplitudes of scalar and tensor perturbations;
  (ii) is it possible to obtain small tensor-to-scalar
  ratio $ r = \mathcal{P}_{h}/\mathcal{P}_{\zeta}$ and, if so,
  what sort of tuning is required for that; (iii) is it possible to generate
  perturbations in a controllable way, i.e., in the regime where the background evolution
  and perturbations are legitimately described within classical field theory and
  weakly coupled quantum theory, respectively ---
  and if so, does this constrain
  the values of observables.

  We emphasize that we design and study our models
    in the Jordan frame where they have fairly simple structure. We could
    equivalently work in the Einstein frame, but then the analysis would
    be more cumbersome. We comment on the Einstein frame counterparts
    of our findings where appropriate.

    An alert reader would anticipate that the spectral tilts
      (both scalar and tensor) are to large extent determined by the
      value of $\mu$ in \eqref{jan24-22-1}; in particular, red tilts occur at
      $\mu > 1$. This is indeed the case, see Sec.~\ref{subsec:perturbations}.
      In this paper we mostly stick to the $\Lambda$CDM value of the scalar
      spectral index~\cite{Planck:2018vyg},
      \begin{equation}
    n_S = 0.9649 \pm 0.0042 \; .
\label{jul17-22-1}
\end{equation}
      We comment, however, that the possible presence of early dark 
      energy makes  the scale invariant
      Harrison--Zeldovich spectrum with $n_S = 1$ consistent with
      observations~\cite{Ye:2021nej,Jiang:2022uyg}. Hence, we also briefly
      consider a bounce model with the flat power spectrum.
    
  This paper is organized as follows. We introduce our class of Horndeski
models and discuss early contracting stage of bouncing universe
in Secs.~\ref{sec:preliminaries}, \ref{sec:solution-powerlaw}.
We derive the properties of linearized scalar and tensor perturbations
generated at that stage in Sec.~\ref{subsec:perturbations}, while in
Sec.~\ref{sec:dilatation} we point out the relation between
the flatness of the spectra and approximate dilatation invariance
of the models. Sections \ref{sec:srong-preliminaries} and \ref{sec:u-bound-gen}
are central in our
relatively
general discussion in Sec.~\ref{preliminary_bounce}:
we observe that there is
tension between the small
value of $r$
(and to lesser extent the red scalar spectral tilt),
on the one hand, and the requirement of the absence of
strong coupling, on the other. We consider this issue at qualitative level
in Sec.~\ref{sec:srong-preliminaries} and proceed to quantitative analysis
in Sec.~\ref{sec:u-bound-gen}. We illustrate this tension
in Sec.~\ref{sec:explicit-bounce}, where we
derive
lower bounds on $r$  in very concrete models,
first
with the $\Lambda$CDM scalar tilt \eqref{jul17-22-1},
and then with $n_S=1$.
We conclude in Sec.~\ref{sec:conclude}.
Appendices A, B, C contain details of more technical character. Appendix D is dedicated to the calculation of a specific form of covariant Lagrangian functions for our model. In Appendix E we proceed with conformal transformation of the metric and show how scale factor and Hubble parameter change after this transformation. Finally, Appendix F contains a concrete example of stable and subluminal evolution, which starts from contraction, then bounce occurs and after that Universe turns into general relativity (GR) kination.

\section{ Horndeski models with power-law contraction}
\label{preliminary_bounce}

\subsection{Preliminaries}
\label{sec:preliminaries}

We 
consider
a subclass of Horndeski theories whose action has the form (in the Jordan frame)
    \begin{align}
      \cal S=&\int d^4x \sqrt{-g}
      \left\{ G_2(\phi, X)-G_3(\phi, X)\Box \phi
      + G_4(\phi,X)R + G_{4X}\big[(\Box \phi)^2 - (\nabla_{\mu}\nabla_{\nu}\phi)^2\big]\right\}\;,
    \label{Hor_L}\\
        X =& -\frac{1}{2}g^{\mu\nu}\partial_{\mu}\phi\partial_{\nu}\phi\;,
    \nonumber
    \end{align}
where 
$\Box \phi = g^{\mu\nu} \nabla_\mu \nabla_\nu \phi$ 
and $(\nabla_{\mu}\nabla_{\nu}\phi)^2 = \nabla_{\mu}\nabla_{\nu}\phi \, \nabla^{\mu}\nabla^{\nu}\phi$, 
$R$ is the Ricci scalar. The metric
signature is $(-,+,+,+)$. Unlike 
the general Horndeski theory,
the Lagrangian  \eqref{Hor_L} involves three arbitrary functions
$G_{2,3,4}$ rather than four.
It is convenient to work in the ADM formalism~ \cite{Kobayashi:2019hrl}.
To this end, the metric is written as
 \begin{equation*}
     ds^2=-N^2 d\hat{t}^2 +  
        \gamma_{ij}\left( dx^i+N^i d\hat{t}\right)\left(dx^j+N^j d\hat{t}\right) \;,
    \end{equation*}
 where  $\gamma_{ij}$ is three-dimensional metric,
 $N$ is the lapse function and  $N_i=\gamma_{ij}N^j$ 
 is the shift  vector.
 We denote the general time variable by $\hat{t}$ and reserve the
 notation $t$ for cosmic time.
 By choosing the unitary gauge (in which $\phi$ depends on $\hat{t}$
 only and has prescribed form $\phi=\phi(\hat{t})$), one rewrites the
action as follows,
 \begin{align}
\label{adm_lagr}
        \mathcal{S} = \int d^4x \sqrt{-g}    \left[ A_2 (\hat{t}, N) + A_3 (\hat{t}, N) K 
        +  A_4 (\hat{t}, N)
        (K^2 - K_{ij}^2) + B_4 (\hat{t}, N) R^{(3)}  \right]\; \text{,}
    \end{align}
where
\begin{equation*}
    A_4(\hat{t},N) = - B_4(\hat{t},N) - N\frac{\partial B_4(\hat{t},N)}{\partial N}\;,
\end{equation*}
$^{(3)} R_{ij}$ is the Ricci tensor made of $\gamma_{ij}$,
   $\sqrt{-g} = N\sqrt{\gamma}$,
    $K= \gamma^{ij}K_{ij}$, $^{(3)} R = \gamma^{ij} \phantom{0}^{(3)} R_{ij}$ and
    \begin{align*}
      K_{ij} &\equiv\frac{1}{2N}\left(\frac{d\gamma_{ij}}{d\hat{t}}
      -\,^{(3)}\nabla_{i}N_{j}-\;^{(3)}\nabla_{j}N_{i}\right) ,
    \end{align*}
    is extrinsic curvature of hypersurfaces $\hat{t}=\mbox{const}$. The relationship between the
    Lagrangian functions in the covariant and ADM formalisms is given
    by~\cite{Gleyzes:2014dya, Gleyzes:2013ooa, Fasiello:2014aqa}
    \be
      G_2 =  A_2 - 2XF_{\phi} \text{,} \;  \;\;\;\;
        G_3 = - 2XF_X - F \text{,}  \;  \;\;\;
        G_4 = B_4 \;\text{,}  
         \label{jan23-22-1}
    \ee
    where $N$ and $X$ are related by
    \be
    N^{-1} d\phi/d\hat{t} = \sqrt{2X}\;,
    \label{jan25-22-30}
    \ee
    and
    \begin{equation}
    \label{F}
        F_X = - \frac{A_3}{\left(2X\right)^{3/2}} - \frac{B_{4\phi}}{X}\; \text{.}
    \end{equation}
We note in passing  a subtlety here. Equation \eqref{F} defines
      $F(\phi, X)$ up to additive term $D(\phi)$. This term modifies the
Lagrangian functions,
\be
G_2 \to  G_2  - 2X D_\phi\;, \;\;\;\;\; G_3 \to  G_3  - D \;.\nonumber
\ee
      However, the additional contribution to the action \eqref{Hor_L} vanishes
      upon integration by parts,
\be
      \int d^4 x \sqrt{-g} (-2X D_\phi + D \Box \phi) =
      \int d^4 x \sqrt{-g} \nabla_\mu(D \nabla^\mu \phi) =0\; .
      \label{jun9-22-1}
\ee
Therefore,
this freedom is, in fact, irrelevant.

To describe FLRW background and perturbations about it, one writes 
\begin{subequations}
  \label{jul17-22-2}
\begin{align}
        N &=N_0(\hat{t}) (1+\alpha)\;, \\
        N_{i} &=\partial_{i}\beta +  N^T_i\;,
        \\
        \gamma_{ij} &=a^{2}(\hat{t}) \Big(\text{e}^{2\zeta}(\text{e}^{h})_{ij} + \partial_i\partial_j Y 
        + \partial_i W^T_j + \partial_j W^T_i\Big) \; ,
\end{align}
\end{subequations}
        where
        $a(\hat{t})$ and $N_0(\hat{t})$ are background solutions,
        $\partial_i N^{Ti}=0$ and
         \begin{align*}
        (\text{e}^h)_{ij} &=\delta_{ij}+h_{ij}+\frac{1}{2}h_{ik}h_{kj}+\frac{1}{6}h_{ik}h_{kl}h_{lj}+
        \cdots\;, \quad h_{ii}=0\;, \quad
        \partial_{i}h_{ij}=0 \; .
    \end{align*}
    Throughout this paper, we  denote the background lapse 
    function by $N$ instead of $N_0$.    
    The residual gauge freedom is fixed
    by setting $Y = 0$ and $W^T_i = 0$.
    Variables
      $\alpha$, $\beta$ and $N^T_i$ enter the action without temporal
    derivatives;
    the dynamical degrees of freedom are $\zeta$
    and transverse traceless $h_{ij}$, i.e., scalar and tensor perturbations.
   Upon integrating out variables $\alpha$ and $\beta$, one obtains
the quadratic actions for scalar and tensor perturbations~\cite{Kobayashi:2015gga}
\begin{subequations}
  \label{jan23-22-5}
  \begin{align}
\label{second_scalar}
\mathcal{S}_{\zeta \zeta}^{(2)}&=\int d\hat{t} d^{3} x  N a^3
\left[
  \frac{\mathcal{G}_S}{N^2}
  \left(\frac{\partial{\zeta}}{\partial \hat{t}}\right)^{2} -
  \frac{\mathcal{F}_S}{a^2} \left (\vec{\nabla} \zeta\right)^{2}\right] \; ,
\\
\mathcal{ S}_{hh}^{(2)}&=\int d\hat{t} d^3x \frac{N a^3}{8}\left[
  \frac{\mathcal{G}_T}{N^2}
            \left (\frac{\partial h_{ij}}{\partial\hat{t}}\right)^2
        -\frac{\mathcal{ F}_T}{a^2}
         h_{ij,k} h_{ij,k}
        \right] \; .
        \label{second_tensor}
  \end{align}
\end{subequations}
Explicit expressions for the coefficients  $\mathcal{G}_S$,  $\mathcal{F}_S$,   $\mathcal{G}_T$,  and
$\mathcal{F}_T$
in general models of the type \eqref{adm_lagr}
as well as equations for background are collected in Appendix A. 

\subsection{Power-law contraction}
\label{sec:solution-powerlaw}
To build a bouncing model with the early-time behavior \eqref{jan24-22-1}, we choose
the following form ~\cite{Ageeva:2021yik} of the Lagrangian functions in \eqref{adm_lagr}
at early times, $t \to -\infty$:
\begin{subequations}
\label{A_old}
	\begin{align}
	&A_2(\hat{t},N) = \hat{g} (-\hat{t})^{-2\mu -2} \cdot a_2 (N) \;\text{,}  \\
	  &A_3 (\hat{t},N)= \hat{g} (-\hat{t})^{-2\mu -1} \cdot a_3 (N)\; \text{,}
          \\
	  A_4 =&A_4 (\hat{t})= -B_4(\hat{t}) = -  \frac{\hat{g}}{2} (-\hat{t})^{-2\mu}\; ,
          \label{A4old}
	\end{align}
\end{subequations}
where $\hat{g}$ is some positive
constant. Then the equations for background, eqs.~\eqref{eoms}, take the
following form
\begin{subequations}
\begin{align*}
 &\frac{(Na_2)_N}{(-\hat{t})^{2}} + \frac{3 N a_{3N} H}{(-\hat{t})} + 3 H^2 =0\;,\\
  &\frac{a_2}{(- \hat{t})^{2}} + 3 H^2 -
      \frac{1}{N}\left[\frac{(2\mu +1) a_3}{(- \hat{t})^2} -
        \frac{4 \mu H}{(-\hat{t})} - 2 \frac{dH}{d\hat{t}} +
        \frac{a_{3N}}{(-\hat{t})} \frac{d N}{d\hat{t}} \right] =0 \; ,
\end{align*}
\end{subequations}
  where $H$ is the physical Hubble parameter.
We make use of the {\it Ansatz}
\be
N=\mbox{const} \; , \;\;\;\;\; a= d (-t)^\chi \; ,
\label{jan31-22-1}
\ee
where $\chi > 0$ is a constant and $t= N\hat{t}$ is cosmic time, so that $H=\chi/t$,
and find the algebraic equations for $N$ and $\chi$:
\begin{subequations}
  \label{jan24-22-10}
  \begin{align}
(Na_2)_N - 3\chi a_{3 N} +3 \frac{\chi^2}{N^2}   &= 0\;,
\label{aug12-21-1}
    \\
    a_2 - \frac{1}{N} (2\mu+1) \Big(a_3 + 2 \frac{\chi}{N} \Big) + 3 \frac{\chi^2}{N^2}   &=0\;.
    \label{aug12-21-2}
  \end{align}
\end{subequations}
In what follows we assume that these equations have a solution with $N>0$ and $1>\chi>0$
(the reason for requiring that $\chi<1$ will become clear shortly, see also
eq.~\eqref{jan24-22-11}).

Let us emphasize that the form of the Lagrangian functions
\eqref{A_old}, as well as the entire discussion in this paper,
refer to the early contraction stage only. To obtain the bounce itself,
as well as subsequent expansion stage, one has to make use of
more sophisticated Lagrangian functions that can be obtained, e.g., by
gluing procedure elaborated in Ref.~\cite{Ageeva:2021yik}. A lesson
from Ref.~\cite{Ageeva:2021yik} is that designing stable bouncing models with
given early-time asymptotics with ``strong gravity in the past''
is relatively straightforward.
In this
paper we do not
aim at constructing complete cosmological models and stick to early times
when the cosmological perturbations are supposed to be generated.

The coefficients entering the quadratic actions
\eqref{jan23-22-5} for perturbations
are straightforwardly calculated by making use of the general
expressions \eqref{eq:Ft_Gt_form}, \eqref{eq:Fs_Gs_form}.
In what follows it is convenient to work in cosmic time $t= N\hat{t}$ and
write the quadratic actions in convenient forms (hereafter dot denotes
the derivative with respect to cosmic time $t$)
\begin{subequations}
  \label{jan21-24-2}
  \begin{align}
      \label{jan24-22-2a}
 \mathcal{ S}_{hh} & =\int dt d^3x \frac{ a^3}{8}\left[
        \mathcal{ G}_T
        \dot h_{ij}^2
        -\frac{\mathcal{ F}_T}{a^2}
        h_{ij,k} h_{ij,k} \right] \; ,
 \\
   \mathcal{ S}_{\zeta\zeta} &=\int dt d^3x a^3\left[
        \mathcal{ G}_S
        \dot\zeta^2
        -\frac{\mathcal{ F}_S}{a^2}
        \zeta_{,i}\zeta_{,i}
        \right] \; .
   \label{feb1-22-3}
\end{align}
\end{subequations}
Then
\be
\mathcal{G}_T= \mathcal{F}_T =  \frac{g}{(-t)^{2\mu}}\;,
\label{jan31-22-2}
\ee
where
\be
g = N^{2\mu} \hat{g}\;,
\label{jul7-22-1}
  \ee
  and
  \be
\mathcal{G}_S=  g \frac{g_S}{2(-t)^{2\mu}} 
\; , \;\;\;\;\; \mathcal{F}_S =  g \frac{f_S}{2(-t)^{2\mu}}\;,
\label{jan31-22-3}
\ee
with
\begin{subequations}
  \label{feb5-22-1}
    
\begin{align}
  f_S &= \frac{2(2-4 \mu + N^2 a_{3N})}{2\chi - N^2 a_{3N}}\;,
 \label{jan25-22-21a}
  \\
    g_S &= 2   \left[\frac{2 \Big(2 N^3 a_{2N}+ N^4 a_{2NN} - 
    3 \chi (2 \chi + N^3 a_{3NN})\Big)}{(N^2
    a_{3N}-2\chi)^2} + 3\right]\;,
    \label{jan25-22-21b}
\end{align}
\end{subequations}
where $a_{2,3}(N)$ and their derivatives are to be evaluated on the solution to
eqs.~\eqref{jan24-22-10}, so that $g$, $f_S$ and $g_S$
are independent of time.
Note that the propagation of the tensor perturbations  is luminal,
\be
 u_T^2 = \frac{\mathcal{F}_T}{\mathcal{G}_T} = 1 \; ,\nonumber
 \ee
 whereas the sound speed in the scalar sector is given by
 \be
 \label{us}
 u_S^2 = \frac{\mathcal{F}_S}{\mathcal{G}_S} = \frac{f_S}{g_S} \; ,
 \ee
 and can be substantially smaller than 1.

 To end up this Section, we notice that the conformal
   transformation
   \be
   \label{conf_transf}
   g_{\mu \nu} = \frac{M_P^2}{2B_4  } g_{\mu \nu \,  (E)}\;, 
   \ee
   where $M_P = (8\pi G)^{-1/2}$ is the reduced Planck mass, converts the
   theory into the Einstein frame. In the Einstein frame, the action
   is cubic Horndeski, and the scale factor is given by
   eq.~\eqref{jan24-22-11}. This justifies the discussion in
   Sec.~\ref{sec:intro} after  eq.~\eqref{jan24-22-11}.

\subsection{Generating perturbations}
\label{subsec:perturbations}

As pointed out in Introduction, the quadratic actions for perturbations coincide
 (modulo the fact that $u_S^2 \neq 1$) with the action for tensor perturbation in
 power-law
 expansion
 setup. So, the cosmological perturbations with nearly flat
 power spectrum may be generated at early contraction epoch. Let us consider for definiteness
 scalar perturbation $\zeta$. It obeys the linearized equation
 \begin{equation}
   \ddot{\zeta}+\frac{2 \mu-3 \chi}{|t|} \dot{\zeta}+\frac{u_{S}^{2} k^{2}}{d^{2} |t|^{2 \chi}} \zeta=0\;,
   \label{jan31-22-5}
 \end{equation}
 For $0<\chi<1$, the mode is effectively
 subhorizon at early times (in the sense that
 the effective Hubble time scale $|t|$ is greater
 than
 the period of oscillations
 $d \cdot |t|^\chi/ (u_S k)$),
 so it is adequately described within the WKB
 approximation. At later times, the mode is superhorizon;
in what follows we consider the case
  \be
  2\mu - 3 \chi >0 \; ,
  \label{jul5-22-100}
  \ee
in which
  the superhorizon mode
 experiences the Hubble friction and freezes out. The horizon exit occurs at
 time $t_f (k)$ when
 \be
 \frac{2\mu - 3\chi}{|t_f|} \sim \frac{u_{S} k}{d |t_f|^{ \chi}}\;,\nonumber
   \ee
   i.e.,
\begin{equation}
\label{t_f}
    |t_f|(k) \sim  \left[\frac{d}{k}\cdot \frac{(2\mu-3\chi)}{u_S}\right]^{\frac{1}{1-\chi}} \; .
\end{equation}
Thus, once the parameters of the theory are chosen in such a way that
$1> \chi >0$, $2\mu - 3 \chi >0$,
 the perturbations are
generated in a straightforward way,
{\it provided that the weak coupling regime occurs
  all the way down to
  $|t| \sim |t_f|$}.
  
Let us also discuss a constraint that has to do with
the Belinsky--Khalatnikov--Lifshitz phenomenon
\cite{Lifshitz:1963ps, Belinsky:1970ew,Belinskii:1972sg}.
In our model this phenomenon manifests itself in the behavior of
superhorizon tensor (and also scalar) modes in the
contracting Universe: in the BKL case, one of the two
solutions for
a superhorizon mode of given conformal momentum grows as $t$ increases
and diverges in the formal limit $t \to 0$, while  another solution
stays constant in time. This means that the Universe becomes strongly
anisotropic and inhomogeneous at late times, which is undesirable
(see, e.g., Ref.~\cite{Erickson:2003zm} for discussion). 
In order to avoid BKL, one makes sure that
the time-dependent superhorizon solution decays, instead of growing,
as $t$ increases towards zero.
In our framework, the equation of motion for
superhorizon perturbation 
  is obtained from \eqref{jan24-22-2a} with
  spatial derivatives neglected,
  \[
  \frac{1}{a^3  \mathcal{G}_T} \frac{d}{dt} \left(a^3  \mathcal{G}_T
  \dot{h}_{ij} \right) = 0 \; .
  \]
  One of its solutions is constant in time, while another is
  \[
  h_{ij} \propto \int~dt~ \frac{1}{a^3  \mathcal{G}_T} \propto
  (-t)^{2\mu -3\chi+1} \, .
  \]
  It decays as $t$ increases towards zero for
  \begin{equation}
  \label{BKL_cond}
    2\mu +1 >  3\chi \;.
  \end{equation}
This constraint ensures also that the BKL phenomenon is absent for
scalar perturbations. Recall that we choose the following constraints on parameters $\mu >1$ and $\chi<1$ in our model, thus the condition \eqref{BKL_cond} is surely satisfied.

We outline the calculation in Appendix B, and here we
give the results for the power spectra:
\begin{equation}
  \mathcal{P}_{\zeta} \equiv \mathcal{A}_{\zeta}\left(\frac{k}{k_*}\right)^{n_S-1}\;,
  \;\;\;\;\;  \mathcal{P}_{T} \equiv \mathcal{A}_{T}\left(\frac{k}{k_*}\right)^{n_T} \; ,
    \label{general_ampl}
\end{equation}
where $k_*$ is pivot scale, the spectral tilts are
\begin{equation}
    n_S - 1 = n_T= 2\cdot \left(\frac{1-\mu}{1-\chi}\right) 
    \label{general_n_s}\;,
\end{equation}
and the amplitudes are given by
  \begin{subequations}
    \label{jan25-22-20}
\begin{align}
\label{amplitude}
\mathcal{A}_{\zeta} &= \frac{1}{g}
\cdot\frac{1}{g_S u_{S}^{2 \nu}} \frac{(1-\chi)^{2 \frac{\mu-\chi}{1-\chi}}}{\pi
  \sin ^{2}(\nu \pi) \Gamma^{2}(1-\nu)}\left(\frac{k_*}{2 d}\right)^{2 \frac{1-\mu}{1-\chi}} \; , \\
\mathcal{A}_{T} &= \frac{8}{g} \cdot \frac{(1-\chi)^{2 \frac{\mu-\chi}{1-\chi}}}{\pi
  \sin ^{2}(\nu \pi) \Gamma^{2}(1-\nu)}\left(\frac{k_*}{2 d}\right)^{2 \frac{1-\mu}{1-\chi}},
\label{a_T}
\end{align}
\end{subequations}
where
\be
\nu = \frac{1+2 \mu-3 \chi}{2(1-\chi)} = \frac{3}{2}
+ \frac{1-n_S}{2}\; .
\label{feb4-22-1}
\ee
We immediately see from eqs.~\eqref{jan25-22-20} that the smallness
of the scalar and tensor amplitudes is guaranteed by the large value of the
overall pre-factor $\hat{g}$ in eqs.~\eqref{A_old},
and hence the factor $g$ given by \eqref{jul7-22-1}.
Also, we see from ~\eqref{jan25-22-20}
that the tensor-to-scalar ratio in our model is
\be
r = \frac{\mathcal{A}_{T}}{\mathcal{A}_{\zeta}} =
8 \frac{f_S^{\nu}}{g_S^{\nu - 1}}
  =  8 g_S u_S^{2\nu}
\; .
\label{feb1-22-2}
\ee
This shows that it is not straightforward to have a
small value of $r$, as required by
observations~\cite{Planck:2018vyg,BICEP:2021xfz,Tristram:2021tvh},
 which give
\be
r<0.032 \; .\nonumber
\ee
Since $f_S\leq g_S$ to avoid superluminality, small $r$ requires
that either  $f_S\leq g_S \ll 1$ or $u_S\ll 1$ or both. 
It is clear from \eqref{feb5-22-1} that obtaining both $g_S \ll 1$
and $f_S \ll 1$ 
requires strong fine-tuning. On the contrary,
eq.~\eqref{jan25-22-21a} suggests that ensuring that  
$f_S\ll 1$ while $g_S \sim 1$,
and hence $u_S^2\ll 1$ may not be so problematic.
We give
concrete examples in Sec.~\ref{sec:explicit-bounce}.
Thus, the small tensor-to-scalar ratio in our set of models is
due to small sound speed of scalar perturbations. This is reminiscent
of the situation in
k-inflation~\cite{Garriga:1999vw,Mukhanov:2005bu,Langlois:2008wt},
  where the tensor-to-scalar ratio is also suppressed for small $u_S$.

We now turn to the scalar and tensor tilts given by
eq.~\eqref{general_n_s}. First, we note that the two tilts are
equal to each other, unlike in most of inflationary
models. Second, we
point out that 
%
approximate flatness of the spectra
is ensured in our set of models by choosing
$\mu \approx 1$, while the slightly red $\Lambda$CDM spectrum
\eqref{jul17-22-1}
is found for
\be
\mu > 1 \; .\nonumber
\ee
As we discuss below,
the small value of $r$, especially
in the case $\mu > 1$, is
non-trivial from the viewpoint
of the strong coupling problem. Before coming to the
strong coupling issue, let us
make a point
on approximate flatness itself.

\subsection{Flatness of power spectra and dilatation invariance}
\label{sec:dilatation}

Flatness of the power spectra at $\mu=1$ is not an accident:
the model with $\mu=1$ is invariant under scale transformations.
Let us see this explicitly.

One immediately observes that
for $\mu = 1$, the ADM action \eqref{adm_lagr} with the Lagrangian functions
given by \eqref{A_old} is invariant under scale transformation
\be
\hat{t} = \lambda \hat{t}^\prime \; , \;\;\;\;\;
x^i = \lambda x^{\prime i} \; , \;\;\;\;\;
(N, N_i, \gamma_{ij} )(x^i, \hat{t}) = (N', N_i^\prime, \gamma_{ij}^\prime)
(x^{\prime i}, \hat{t}^{\prime})\;, 
\label{jan25-22-60}
\ee
with $\lambda = \mbox{const}$.
However, in the ADM language this is a somewhat murky point. To clarify it,
we move to covariant formalism with the action \eqref{Hor_L}. To this end,
we
define the field $\phi$, without loss of generality, in such a way that
\begin{equation*}
    -\hat{t} = \text{e}^{-\phi} \; .
\end{equation*}
Then eq.~\eqref{jan25-22-30} gives
\begin{equation}
    N = \frac{\text{e}^{\phi}}{\sqrt{2X}} \;,
\label{jan25-22-40}
\end{equation}
and the Lagrangian functions take the following forms
%
%
    \begin{subequations}
    \label{A_phi_X}
    \begin{align}
      &A_2 =  \hat{g}
      \text{e}^{(2\mu+2)\phi} a_2\Big(\frac{\text{e}^{\phi}}{\sqrt{2X}}\Big) \;\text{,} \\ 
        &A_3 = \hat{g} \text{e}^{(2\mu+1)\phi} a_3\Big(\frac{\text{e}^{\phi}}{\sqrt{2X}}\Big)\; \text{,} \\
      &A_4 =  - \frac{\hat{g}}{2}
      \text{e}^{2\mu\phi}\; .
    \end{align}
    \end{subequations}
    They define the Lagrangian functions $G_2$, $G_3$, $G_4$ in accordance with
    eqs.~\eqref{jan23-22-1} and \eqref{F}. We put full expressions for these functions $G_2$, $G_3$, $G_4$ and some helpful calculations in Appendix D.

    Now, in covariant formalism we introduce the scale transformation of metric
    and scalar field,
    \be
    g_{\mu \nu} = \lambda^2 g_{\mu \nu}^\prime \; , \;\;\;\;
    \phi = \phi^\prime - \ln \lambda\;,
\label{jan25-22-70}
    \ee
    so that $X= \lambda^{-2} X^\prime$.
    The combination in the right hand
    side of eq.~\eqref{jan25-22-40} is invariant under this transformation.
    The meaning of this property is that $N^\prime =
    \frac{\text{e}^{\phi'}}{\sqrt{2X'}}$
    (which is actually equal to $N$)
    is the lapse function in coordinates $x^{\prime \mu}$ introduced in
    \eqref{jan25-22-60}.  With this understanding,
    it is fairly straightforward to check that the action \eqref{Hor_L}
    with $\mu=1$
    is invariant under scale transformation \eqref{jan25-22-70}. This
    is clear from the fact that under scale transformation one
    has
    \be
    A_2 (\phi, X) =  \lambda^{-2\mu-2} A_2 (\phi', X')\; ,
    \; \;\; A_3 (\phi, X) =  \lambda^{-2\mu-1} A_3 (\phi', X') \; ,
    \;\;\; B_4 (\phi) =  \lambda^{-2\mu} B_4 (\phi')\;,\nonumber\ee
    and 
     $\Box \phi =     \lambda^{-2} \Box^\prime \phi'$, 
    $R = \lambda^{-2} R'$.
    A subtlety here concerns
      the function $F$. Its derivative $F_X$
    transforms as $F_X = \lambda^{2-2\mu} F^\prime_{X^\prime}$, as it should,
    so that one would think that
    \be
    F = \int F_X dX = \lambda^{-2\mu} \int  F^\prime_{X^\prime} dX' =
    \lambda^{-2\mu}   F^\prime\; .\nonumber
    \ee
    This is not quite true, though.
    $F_X$ defined by eq.~\eqref{F} may contain a term
    $c\mbox{e}^{2\mu \phi} X^{-1}$,
    i.e., $F$ may contain a term $c \mbox{e}^{2\mu \phi} \ln X$.
    Then, upon scaling
    transformation, function $F$,
    and hence
    functions $G_2$ and $G_3$ obtain log-$\lambda$ terms,
    \be
    (G_2)_{log} = -4 c \mu \ X^\prime \mbox{e}^{2\mu \phi'}
    \cdot \lambda^{-2\mu -2} 
        \ln \lambda^{-2}\; , \;\;\;\;\;
    (G_3)_{log} = - c \ \mbox{e}^{2\mu \phi'}\cdot \lambda^{-2\mu}
    \ln \lambda^{-2}\; .\nonumber
    \ee
    However, their contribution to the action \eqref{Hor_L} vanishes upon
    integration by parts,  in the same way as in
      eq.~\eqref{jun9-22-1}. Modulo this subtlety, the functions
    $G_2$, $G_3$, $G_4$ defined by 
    eqs.~\eqref{jan23-22-1} and \eqref{F}, have correct scaling at $\mu=1$
    which ensures the scale invariance of the theory in covariant formulation.

    \subsection{Tension between small $r$ and
      strong coupling: preliminaries}
\label{sec:srong-preliminaries}

In this Section, we discuss
  in general terms the problematic issue
  with our mechanism of the generation of the cosmological perturbations
 at
Horndeski bounce. It has to do with the dangerous strong coupling, on the one
hand,
and the small value of $r$, on the other --- especially for
positive $(\mu -1)$ as required for
the red scalar tilt.
Using a concrete example,
we will see in Sec.~\ref{sec:explicit-bounce} that the problem may be overcome,
but in a quite narrow range of parameters. This makes our mechanism
particularly interesting and falsifiable.

In this Section
we mostly consider for definiteness
the case $\mu > 1$,  as required by the $\Lambda$CDM value of the
scalar spectral index \eqref{jul17-22-1},
see eq.~\eqref{amplitude}. Our formulas, however,
are valid also in the Harrison--Zeldovich case $\mu=1$, $n_S=1$.
We make specific
comments on the latter case in appropriate places.

Taken literally,
the model with $\mu>1$
suffers strong coupling problem in the asymptotic past,
$t \to -\infty$. This has been discussed in detail
in Ref.~\cite{Ageeva:2021yik} (see also
Refs.~\cite{Ageeva:2018lko,Ageeva:2020gti,Ageeva:2020buc}); here we
sketch the argument.

The characteristic classical energy scale in the power-law bounce model
is the inverse
time scale of evolution,
\be
E_{class} (t) = |t|^{-1} \; .\nonumber
\ee
Indeed, both background values of physical quantities and parameters governing the
perturbations evolve in power-law manner and get order 1 changes in time interval
of order $|t|$ (as an exception,
this does not apply to the scale factor $a(t)$ for $\chi \ll 1$, but
does apply to the Hubble parameter, since $|\dot{H}/H| \sim |t|^{-1}$). To see whether this
classical energy scale is lower than the quantum strong coupling scale, one has to estimate the
latter.

Let us consider first
the tensor sector of the model. Its quadratic action is given
by eq.~\eqref{jan24-22-2a}; importantly, the coefficient
${\mathcal{ G}_T}={\mathcal{ F}_T}$ tends to zero as $t \to - \infty$, 
see \eqref{jan31-22-2}. 
The cubic action reads~ \cite{Gao:2011vs}
   \begin{eqnarray}
        \mathcal{S}^{(3)}_{hhh}
        =  \int dt\text{ }a^3d^3x\Big[\frac{\mathcal{ F}_T}{4a^2}\left(h_{ik}h_{jl}
        -\frac{1}{2}h_{ij}h_{kl}\right)h_{ij,kl} 
        \Big] \; .
\label{jun11-22-30}
   \end{eqnarray}
   Thus, the quantum strong coupling energy scale $E_{strong}$
   in the tensor sector
   is determined by the behavior of ${\mathcal{ F}_T}$.
   To estimate this scale at a given moment of time, we note first that we can
   rescale spatial coordinates at that moment
   of time to set
   \be
   a=1 \; .\nonumber
   \ee
   Now, if the strong coupling scale  $E_{strong}$ is much higher than
   the energy scale
$|t|^{-1}$ of the classical evolution, the
   background can be treated as slowly varying, and at a given moment of time
   it is natural to introduce canonically normalized field $h_{ij}^{(c)}$ by
   \be
   h_{ij} = {\mathcal{G}_T}^{-1/2}  h_{ij}^{(c)} \; .\nonumber
   \ee
   Then the cubic interaction term becomes
   \be
   \mathcal{S}^{(3)}_{hhh} 
   =\int d t \text{ }d^3x
   \Big[\frac{\mathcal{ F}_T}{4\mathcal{G}_T^{3/2}}\left(h^{(c)}_{ik}h^{(c)}_{jl}
     -\frac{1}{2}h^{(c)}_{ij}h^{(c)}_{kl}\right)
     \partial_k \partial_l h^{(c)}_{ij}
        \Big]\; .\nonumber
   \ee
   On dimensional grounds, the quantum strong coupling scale is estimated
   as
   \be
   \label{ETTT}
   E_{strong}^{TTT} \sim \frac{\mathcal{G}_T^{3/2}}{\mathcal{ F}_T} =
   \frac{g^{1/2}}{|t|^\mu} \; ,
   \ee
   where we use  \eqref{jan31-22-2}. This scale is indeed higher than the
   classical  energy scale $H\sim |t|^{-1}$ provided that
   \be
|t|^{2\mu-2} < g \; .
   \label{feb1-22-1}
   \ee
   As pointed out in
   Refs.~\cite{Ageeva:2018lko,Ageeva:2020gti,Ageeva:2020buc,Ageeva:2021yik},
   this inequality is indeed valid at arbitrarily large $|t|$
   for $\mu < 1$, but {\it it does not hold in the asymptotic past}
   for $\mu > 1$, as required for the red spectral tilt.

   Thus, there is potential
   tension between the red tilt and  the validity of the
   (semi-)classical field theory treatment, i.e.,
   absence of strong coupling. One  may
   take various attitudes towards this
   potential problem. First, one  may
   pretend to be ignorant about
   the situation at very early times, and consider only the evolution
   at the epoch when the theory is weakly coupled, in the sense that
   $|t|^{-1} < E_{strong}$. Second, one 
   may
   think of a slow change of the
   exponent $\mu = \mu(t)$ from $\mu<1$ in the asymptotic past
   to $\mu > 1$ at later times, when the perturbations are generated.
   In any case, however, our calculation of
   the power spectra in Sec.~\ref{subsec:perturbations} and Appendix B
   is valid  {\it provided that the WKB evolution before
     the exit from effective horizon occurs in the weak coupling regime}.
    This means
   that the freeze-out time \eqref{t_f} must obey the weak
   coupling condition  \eqref{feb1-22-1} for any relevant momentum
   $k$.

   In fact, the tensor sector is not problematic in this
     regard. To see this, we recall that the tensor modes exit the effective
     horizon at
     \be
     t_f^{(T)} (k) \sim \left(\frac{d}{k}\right)^{\frac{1}{1-\chi}},\nonumber
     \ee
     (see eq.~\eqref{t_f} with $u_S = 1$ for tensor modes). Then the
     relation~\eqref{feb1-22-1} with $ t=  t_f^{(T)}$ becomes
\be
     \frac{1}{g} \left(\frac{d}{k}\right)^{2\frac{\mu-1}{1-\chi}} \ll 1.\nonumber
     \ee
     The left hand side here is of the order of the tensor amplitude
     ${\cal A}_T$, eq.~\eqref{a_T}, so that the absence of strong
     coupling at the horizon exit time is guaranteed by the smallness of
     the tensor amplitude.

   The latter property is actually obvious from
     the Einstein frame viewpoint.
     For $\mu > 1$, the Einstein frame universe experiences the
     power-law inflation~\eqref{jun23-22-1}. Strong coupling in the asymptotics
     $t\to -\infty$ ($t_E \to 0$) is interpreted as a mere fact that
     the inflationary Hubble parameter $H_E \sim t_E^{-1}$ formally exceeds
     $M_P$ at small $t_E$. Now, the tensor amplitude is of order $H_E^2/M_P^2$
     at the exit time from the inflationary horizon; small tensor amplitude
     means the absence of strong coupling at that time, $H_E \ll M_P$.

    In the case  $\mu = 1$,   the condition of
       validity of the classical description is time-independent,
\begin{equation*}
    g\gg1 \; .
\end{equation*}
Again, this condition is automatically satisfied provided that
the tensor amplitude \eqref{a_T} is small.

 The situation is  more
 subtle in the scalar sector, 
   since
   the scalar sound speed $u_S$ is small, as required by
   the small tensor-to-scalar ratio (see eq.~\eqref{feb1-22-2}).
   To appreciate this  new aspect,
   we consider
   scalar perturbations whose quadratic action is given by
   \eqref{feb1-22-3}, i.e.,
   \be
    \mathcal{ S}_{\zeta\zeta} =\int dt d^3x a^3   \mathcal{ G}_S \left[
        \dot\zeta^2
        -\frac{u_S^2}{a^2}
        \zeta_{,i}\zeta_{,i}
        \right] \; .\nonumber
    \ee
    Hereafter we assume, in view of the above discussion,
    that $g_S$ in eqs.~\eqref{jan31-22-3}, \eqref{jan25-22-21b}
    is of order 1, and the smallness of $u_S$ is due to small $f_S$.
    Cubic terms in the action for $\zeta$ are calculated in
    Refs.~\cite{DeFelice:2011zh,Gao:2011qe,DeFelice:2011uc,Gao:2012ib}.
 As we discuss in  Sec.~\ref{sec:hierarchy}
   and Appendix C,
    the most relevant terms
    reduce to 
    just one term \eqref{jun11-22-22}
    in the cubic action (with $a=1$, as before):
  \begin{align}
        \mathcal{S}^{(3)}_{\zeta\zeta\zeta} 
        =  \int d {t}~d^3 {x} 
        \Lambda_\zeta
               {\partial}^2 \zeta \left( {\partial}_i \zeta \right)^2, 
\label{jun11-22-22a}
   \end{align}
   with
   \be
   \Lambda_\zeta = 
   \frac{\mathcal{ G}_T^3}{4\Theta^2} \; ,\nonumber
   \ee
      where $\partial^2 = \partial_i \partial_i$,
        and $\Theta$ is given by
        \eqref{theta}. In our model \eqref{A_old}, we have
   \be
        \Theta = g \frac{\vartheta}{|t|^{2\mu +1}} \; , \;\;\;\;\;\;
  \vartheta = \frac{1}{2} N^2 a_{3N} - \chi\;, 
\label{jun11-22-15}
  \ee
        Thus,
        \be
        \Lambda_\zeta = g \frac{\lambda_\zeta}{|t|^{2\mu - 2}} \; ,\nonumber
          \ee
             where\footnote{In the
                 model of Sec.~\ref{sec:explicit-bounce}
               the property   $\lambda_\zeta = O(1)$ is valid provided that
                $\chi$ is not fine tuned to be
             very close to 1, which is the case we consider.}
             \be
             \lambda_\zeta = \frac{1}{4\vartheta^2} = O(1)\;,
\label{jun12-22-10}
             \ee
for all values of $u_S$ including $u_S \ll 1$.
%
             To get rid of the sound speed in the quadratic part of the action,
             we not only set $a=1$, but
             rescale the spatial coordinates further,
              $x^i = u_S y^i$. Upon introducing canonically normalized field
             \begin{equation*}
    \zeta^{(c)} = (2\mathcal{G}_S)^{1/2} u_S^{3/2} \zeta\;,
             \end{equation*}
             we obtain the quadratic action
             in canonical form (with effective
             sound speed equal to 1),
             whereas the
             cubic action becomes
             \be
               \mathcal{S}^{(3)}_{\zeta\zeta\zeta} = \int dt d^3y
               \Lambda_\zeta  (2\mathcal{G}_S)^{-3/2} u_S^{-11/2}
              \partial^2_y \zeta^{(c)} (\partial_y\zeta^{(c)})^2
                \; .
\label{jun10-22-1}
      \ee
      On dimensional grounds, the strong coupling energy scale is determined by
      \be
      (E_{strong}^{\zeta \zeta \zeta} )^{-3} \sim   \Lambda_\zeta
      (\mathcal{G}_S)^{-3/2} u_S^{-11/2}\;.\nonumber
      \ee
      Collecting all factors, and omitting  factors
      of order 1, we get
      \be
      E_{strong}^{\zeta \zeta \zeta} \sim \frac{1}{|t|} \left( \frac{g^{1/2} 
      u_S^{11/2}}{|t|^{\mu -1}} \right)^{1/3}.
\label{jun11-22-50}
      \ee
     The condition of validity of (semi-)classical approximation, $E_{strong} > |t|^{-1}$, now reads   
     \be
     \left( \frac{g u_S^{11}}{|t|^{2(\mu -1)}} \right)^{1/6} > 1 \; .
     \label{feb3-22-1}
     \ee
     For small $u_S$ it is stronger than the condition \eqref{feb1-22-1}, i.e., eq.~\eqref{feb3-22-1}
     is valid at
     later times (smaller $|t|$) than eq.~\eqref{feb1-22-1}.

     Let us see whether the condition \eqref{feb3-22-1} can be satisfied
     at the times when the
     relevant modes of perturbations exit the effective horizon.
     The most dangerous are the
     longest  modes, i.e., the smallest $k=k_{min}$. To
       obtain a rough estimate,
     we take $k_{min}\approx k_*$ (the momentum dependence is weak in view of
     small
     $|n_S - 1|$), and relate the exit time  \eqref{t_f} at $k=k_*$
     with the scalar amplitude~\eqref{amplitude}.
     We omit factors of order 1 and obtain
     \be
     t_f^{2(\mu - 1)} \sim g {\cal A}_\zeta u_S^3 \; .\nonumber
     \ee
In this way we find
%
     \be
   \left( \frac{g u_S^{11}}{|t_f(k_{min})|^{2(\mu -1)}} \right)^{1/6}
      \sim  \left( \frac{u_S^{8}}{\mathcal{A}_{\zeta}}  \right)^{1/6}
      \sim  \left( \frac{r^{4/\nu}}{\mathcal{A}_{\zeta}} \right)^{1/6} \; ,\nonumber
     \ee
     where we make use of eq.~\eqref{feb1-22-2} with $\nu$ given
     by \eqref{feb4-22-1}.
     So, the validity condition  \eqref{feb3-22-1}
     for our weak coupling calculations is roughly
\be
     \left( \frac{r^{4/\nu}}{\mathcal{A}_{\zeta}} \right)^{1/6} > 1 \; .
     \label{mar22-22-1}
     \ee
     We see that there is an interplay between two small numbers,
     $r$ and $\mathcal{A}_{\zeta}$. For a crude estimate,
     we take $\chi \ll 1$ and $\mu \approx 1$,
     consistent with small $(1-n_S)$ as given by \eqref{general_n_s}. Then
     $\nu \approx 3/2$. If we then take, as an example, 
$r=0.02$ and insert $\mathcal{A}_{\zeta} \simeq 2\cdot 10^{-9}$ into the left
     hand side of eq.~\eqref{mar22-22-1}, we obtain its numerical value
 approximately equal
     to 5, suspiciously close to 1. The lesson we
     learn from this back-of-envelope estimate is twofold.
     First, one cannot neglect numerical
     factors ``of order 1'' here. In particular, one has to
     be more precise when evaluating the
     strong coupling scale $E_{strong}$: instead of
     naive dimensional analysis, one has to consider 
     unitarity bounds. We
    study this point in general terms in
     Sec.~\ref{sec:u-bound-gen}  and apply the results to 
 a concrete model
     in Sec.~\ref{sec:explicit-bounce}. Second,
     it is clear that one cannot have arbitrarily
     small tensor-to-scalar ratio $r$ in our class
     of models; indeed, $r\simeq 0.02$ appears to be already
     on the edge of the validity of the
     weakly coupled description that we make use of.
     We substantiate the latter observation in
      Sec.~\ref{sec:explicit-bounce} within the concrete model.

      The above analysis goes through also in the case $\mu=1$,
        $n_S=1$. Instead of \eqref{feb3-22-1},
        we 
        obtain the condition for the absence of strong coupling, which is
        again time-independent,
\begin{equation}
\label{mu_1_NOSC}
      \left( g 
      u_S^{11} \right)^{1/6}>1 \; .
\end{equation}
With $\nu = 3/2$ this gives
\begin{equation*}
    \left( \frac{r^{8/3}}{\mathcal{A}_{\zeta}} \right)^{1/6} > 1\; ,
\end{equation*}
which is similar to \eqref{mar22-22-1}.
We refine this qualitative argument in Sec.~\ref{sec:explicit-bounce}.

      \subsection{Tree level unitarity and strong coupling energy scale}
\label{sec:u-bound-gen}

\subsubsection{Unitarity relations with different sound speeds}
\label{sec:speeds-unitarity}

    The quantum energy scale of strong coupling is conveniently evaluated by
    making use of the unitarity bounds on partial wave amplitudes (PWAs)
    of
    $2\to 2$ scattering~\cite{Oller:2019opk,oller.190503.1,Lacour:2009ej,Gulmez:2016scm}.
    In our model we have nine $2 \to 2$ channels, which we collectively
    denote by $\alpha \beta$, where $\alpha = (\alpha 1, \alpha 2)$
    and $\beta = (\beta 1, \beta 2)$
    refer to initial state and final state, respectively:
\begin{subequations}
    \begin{align}
      \alpha 1, \alpha 2 \to \beta 1 , \beta 2 ~
      =~ &\zeta \zeta \to \zeta \zeta\ ,
      \label{jun10-22-2a}\\
      & \zeta h \to \zeta \zeta  \; , \;\;\;\; \zeta \zeta \to \zeta h
     \label{jun10-22-2c} \\
      & \zeta  h  \to \zeta h 
      \\
      &\zeta \zeta \to hh \; , \;\;\;\; hh \to \zeta \zeta
      \label{jun10-22-2b}\\
      & \zeta h \to hh  \; , \;\;\;\; hh \to \zeta h\\
      & hh \to hh \; .
    \end{align}
    \end{subequations}
An additional complication is that the perturbations $\zeta$ and $h$
have different sound speeds.

In this situation a (fairly obvious) generalization of the PWA unitarity
relation is~\cite{Ageeva:2022nbw}
\be
\mbox{Im}~ a^{(l)}_{\alpha \beta} = \sum_\gamma a^{(l)}_{\alpha \gamma}
  \frac{g_\gamma}{u_{\gamma 1} u_{\gamma 2}(u_{\gamma 1}+ u_{\gamma 2})}
  a^{(l) \, *}_{\gamma \beta} \; ,
\label{jun9-22-2}
  \ee
  where
  $ a^{(l)}_{\alpha \beta}$ is PWA with angular momentum $l$ and initial
  and final states $\alpha$ and $\beta$, respectively,
  $\gamma$ refers to two particles in the intermediate state
  with sound speeds $u_{\gamma 1}$ and $u_{\gamma 2}$, and $g_\gamma =2$ if these
  intermediate particles are
  distinguishable and $g_\gamma =1$ if these particles are
  identical.\footnote{Equation \eqref{jun9-22-2} is not the most general
    unitarity relation, but it is valid if the right hand side
    is saturated by the tree level amplitudes. This is sufficient for our
    purposes.} We omitted contributions to the right hand side
  due to multiparticle intermediate states, since
  they can only strengthen the unitarity bound.

  Upon redefining
  \be
  \tilde{a}^{(l)}_{\alpha \beta} =
  \left(\frac{g_\alpha }{u_{\alpha 1} u_{\alpha 2}(u_{\alpha 1}
    + u_{\alpha 2})}\right)^{1/2}
  a^{(l)}_{\alpha \beta} \left(\frac{g_\beta}{u_{\beta 1} u_{\beta 2}
    (u_{\beta 1}+ u_{\beta 2})}\right)^{1/2} \; ,
\label{jun9-22-5}
  \ee
  we arrive at familiar form of the unitarity relation which we write
in the matrix form 
  for
  the matrix $  \tilde{a}^{(l)}_{\alpha \beta}$:
  \be
  \mbox{Im}~ \tilde{a}^{(l)} =  \tilde{a}^{(l)}  \tilde{a}^{(l)\, \dagger} \;.\nonumber
  \ee
  The most stringent tree level
  unitarity bound is obtained for the largest
  eigenvalue of the tree level
  matrix $\tilde{a}^{(l)}$ (which is real and symmetric). This bound
  reads~\cite{Grojean:2007zz}
  \be
  |\mbox{maximum~eigenvalue~of}~ \tilde{a}^{(l)}| \leq \frac{1}{2} \; .\nonumber
  \ee
All these properties are derived in detail in Ref.~\cite{Ageeva:2022nbw}.

\subsubsection{Dimensional analysis for $u_S \ll 1$.}

The model we consider has the large parameter $u_S^{-1}$ which governs
small tensor-to-scalar ratio. So, as we have seen, the earliest time
after which we can trust our setup depends on $u_S$. Let us see that
the dependence
of the rescaled amplitudes $\tilde{a}^{(l)}_{\alpha \beta}$ on $u_S$ is the
only source of
refinement
of the naive estimate
for the time $t_{cl}$ of the onset of the classical theory
(cf.
\eqref{feb1-22-1})
\be
|t_{cl}|^{2\mu -2} \sim g \; .
\label{jun10-22-10}
\ee
We restrict ourselves to the cubic order in perturbations; by experience,
higher orders are expected not to give anything
new~\cite{Ageeva:2020buc,Ageeva:2021yik}.
%
   {\it
Let us ignore the fact that $u_S \ll 1$ for the time being.}
Then the entire Lagrangian defined by the functions \eqref{A_old}
is proportional to $g (-t)^{-2\mu}$, while the only other ``parameter''
is $t$ (we ignore constants of order 1).
Note that $g (-t)^{-2\mu}$
has dimension $(\mbox{mass})^2$. So, on dimensional grounds,
before rescaling to canonically
normalized fields, the terms in the cubic Lagrangian have the following
schematic form
\be
\frac{g}{|t|^{2\mu}}  \cdot (|t|\partial)^n \cdot  \partial^2 \cdot\varphi^3 \;, \nonumber
\ee
where $\varphi$ stands collectively for (dimensionless)
scalar and tensor perturbations, and, with slight abuse of notations,
we do not distinguish
temporal and spatial derivatives
at this
stage. 
Going to canonically normalized fields
$\varphi^{(c)} \sim (g^{1/2}/|t|^{\mu})  \varphi$, we write the cubic Lagrangian
as follows
\be
\frac{|t|^\mu}{g^{1/2}}  \cdot (|t|\partial)^n \cdot\partial^2 \cdot \varphi^{(c) \, 3} \; .
\label{jun11-22-44}
\ee
With this cubic coupling, its contribution to
the $2\to 2$ amplitude is, schematically,
\be
a^{(l)} \sim \frac{\left( (|t|^\mu/g^{1/2}) E^2 (E|t|)^n\right)^2}{E^2}
\sim \frac{|t|^{2\mu -2}}{g} (E|t|)^{2n + 2} \; ,\nonumber
\ee
 where $E$ is the center-of-mass energy, and $E^2$ in
 the denominator comes from the propagator, see Fig.~\ref{fig:diagram}.
This reiterates that ignoring the fact that $u_S \ll 1$, one
would obtain the estimate \eqref{jun10-22-10}
for the time of the onset of the classical theory irrespectively of the
channel considered: at that time the amplitude at energy scale
$E \sim E_{class} =|t|^{-1}$ saturates the unitarity bound.

   {\it
     Let us now reintroduce $u_S \ll 1$.}
   Importantly,
  the coefficients in the cubic
Lagrangian do not contain inverse powers of $u_S$, so, no enhancement
by $u_S^{-1}$ occurs due to the cubic Lagrangian
itself.
Note that this is not entirely trivial. First,
the theory involves 
non-dynamical variables
$\alpha$, $\beta$, $N_i^T$
entering \eqref{jul17-22-2}. Their expressions
  $\alpha (\zeta, h_{ij})$, $\beta(\zeta, h_{ij})$ and $N_i^T (\zeta, h_{ij}))$,
    obtained by solving the constraint equations, may in principle be enhanced
    by inverse powers of $u_S$. As a matter of fact,
    one can check that this is not the case.
    Second, one may be tempted to use linearized equations of motion
    when obtaining the cubic action. This would introduce spurious
    inverse powers of $u_S$ when inserting $\partial_i \partial_i \zeta = u_S^{-2}
    \ddot \zeta$. The latter subtlety is taken care of by working
    consistently off-shell, as we do in what follows.

   Still,
   the rescaled amplitudes
$\tilde{a}^{(l)}$ acquire the dependence on $u_S$. Schematically,
the rescaled amplitudes are now
\be
\tilde{a}^{(l)} \sim \frac{|t|^{2\mu -2}}{g} (E|t|)^{2n + 2} u_S^{-K},\nonumber
\ee
where $K$ depends on the process. Now the time of the
onset of the classical theory is determined by
\be
|t_{cl}|^{2\mu -2} \sim g u_S^K \;.\nonumber
\ee
The larger $K$, the smaller $|t_{cl}|$, the later
    the system
enters the
classical theory/weak coupling regime. So, to figure out the actual time
$t_{cl}$ (the latest of the ``strong coupling times''),
we are going to hunt for processes whose
rescaled amplitudes are enhanced by $u_S^{-K}$ with
the largest value of $K$.

  In the case $\mu=1$, the validity condition for (semi-)classical treatment
  is
  $gu_S^K >1$, so, again, the strongest bound on the paramenetrs of a model
  is obtained for the largest value of $K$.

\subsubsection{Hierarchy of rescaled amplitudes}
\label{sec:hierarchy}

\begin{figure}
    \centering
    \includegraphics[width=0.8\textwidth]{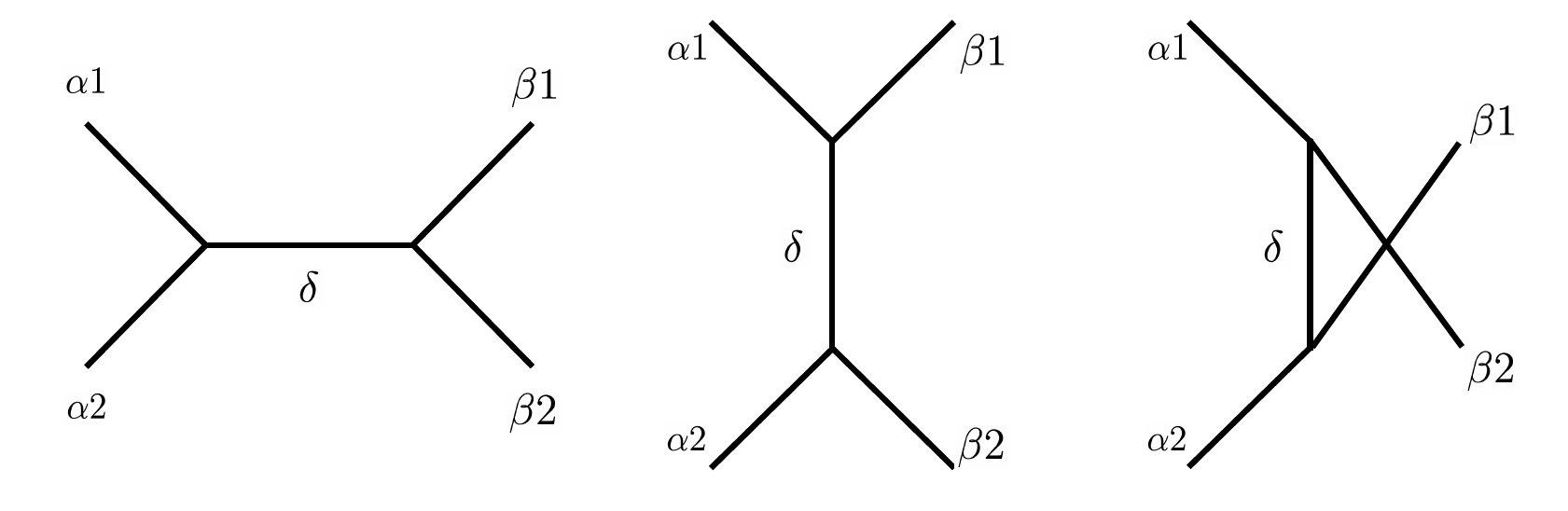}
    \caption{Tree-level diagrams. Particles $\alpha 1$, $\alpha 2$, $\beta 1$, 
    $\beta 2$ and $\delta$ can be scalar or tensor.}
    \label{fig:diagram}
\end{figure}

Let us consider tree-level
diagrams of the types shown in Fig.~\ref{fig:diagram}.
%
%
%

In our class of models, the amplitudes $\tilde{a}^{(l)}_{\alpha \beta}$
with one and the same energy $E$  in the center-of-mass frame
${\bf p}_{\alpha 1} = -{\bf p}_{\alpha 2}$ 
and different particles in the initial, final and
intermediate states
show hierarchical pattern 
in terms of the large parameter
$u_S^{-1}$.
This pattern  is due to the following properties:

(i) Due solely to rescaling,
the rescaled amplitudes \eqref{jun9-22-5} are enhanced by
a factor $u_S^{-3/2}$ for two initial (or two final)
{\it scalar} external legs; by a factor $u_S^{-1/2}$ if initial
(or final) legs are $\zeta h$;
no enhancement of
this sort is associated with two tensor initial (or final)
external legs.

(ii) Since the energy and momentum of a {\it scalar} are related by
$\omega = u_S p$ (we reserve the notation $E$ for the center-of-mass energy),
spatial momentum of an incoming or outgoing scalar may be  either of
order  $p \sim E/u_S$ or  of order $p \sim E$. In the former case (only!)
every {\it spatial} derivative in a vertex, that acts
on external leg  $\zeta$
 gives enhancement
$u_S^{-1}$. The same observation applies to internal line $\zeta$ in $t$-
and $u$-channels,
if spatial momentum transfer is of order $E/u_S$. 

(iii) The scalar propagator is given by
\be
S(\omega,p) = \frac{1}{\omega^2 - u_S^2 p^2} \; .\nonumber
\ee
For $\omega=0$ ($t$-
and/or $u$-channel diagrams with internal line $\zeta$)
this gives enhancement $u_S^{-2}$, provided that the momentum transfer is
$p \sim E$ (but not $E/u_S$).

To proceed further, we note that the maximum number of {\it spatial}
derivatives in triple-$\zeta$ vertex is 4.
In the particular class of Horndeski models \eqref{Hor_L}
with $G_5=0$,
and, furthermore, with $G_4 = G_4 (\phi)$,
there are at most 2 {\it spatial} derivatives in other vertices.
We discuss this point in Appendix C.
Another useful observation is that for a given center-of-mass
energy $E$, incoming (outgoing) momenta are of order
$p\sim E/u_S$ if {\it both} initial (final) particles are $\zeta$, and
$p\sim E$ otherwise.

\begin{figure}
    \centering
    \includegraphics[width=0.8\textwidth]{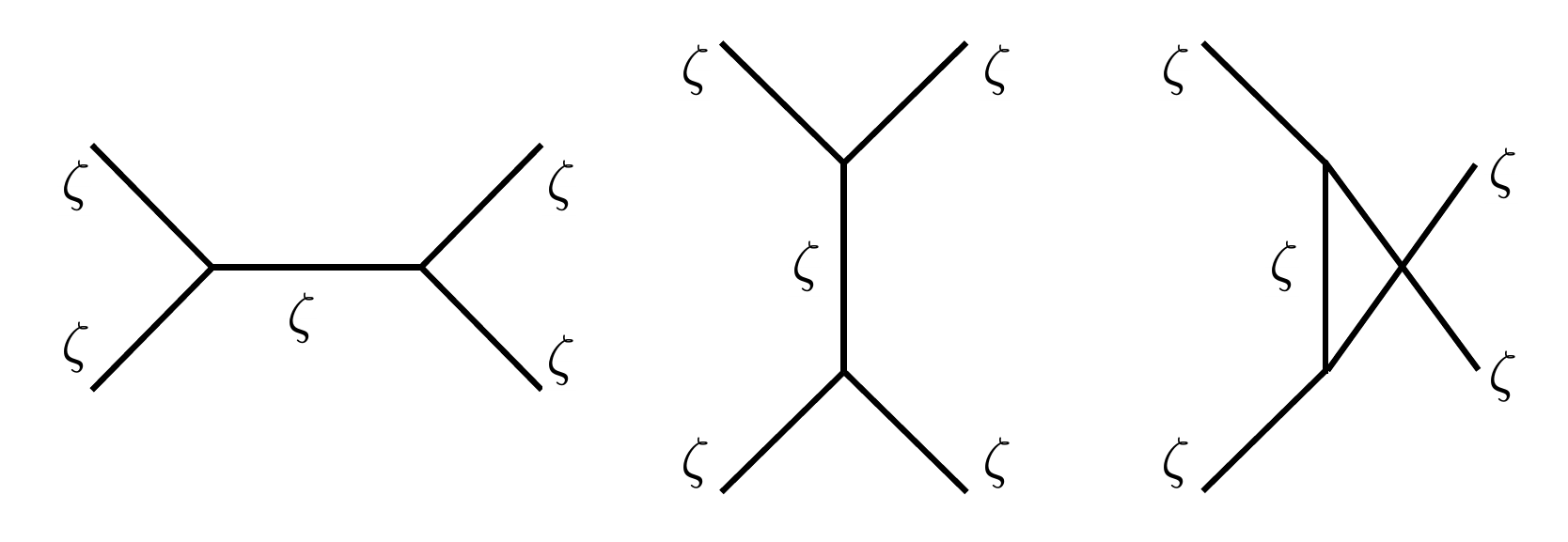}
    \caption{Purely scalar tree-level diagrams:
      case $\textbf{(a)}$.}
    \label{fig:diagr_3scalar}
\end{figure}

We now consider various channels and diagrams separately.

{\bf (a)}. Purely scalar diagrams,
Fig.~\ref{fig:diagr_3scalar}, process \eqref{jun10-22-2a}.
In this case all spatial momenta, including intermediate momentum
in $t$- and $u$-channel diagrams, are of order $E/u_S$. Hence, the enhancement
mechanisms (i) and (ii) are at work, while the mechanism (iii) is not.
The maximum number of spatial derivatives at each vertex is 4,
so the diagrams are of order
\be
u_S^{-3/2} \cdot u_S^{-3/2} \cdot 1 \cdot u_S^{-4} \cdot  u_S^{-4} = u_S^{-11}, \; \nonumber
\ee
(hereafter the first two factors are
   due to enhancement (i),
the third factor   due to enhancement (iii) and the last two factors
 due to enhancement (ii)).
This precisely matches the amplitude that one obtains from the
cubic action \eqref{jun10-22-1}.

\begin{figure}
    \centering
    \includegraphics[width=0.8\textwidth]{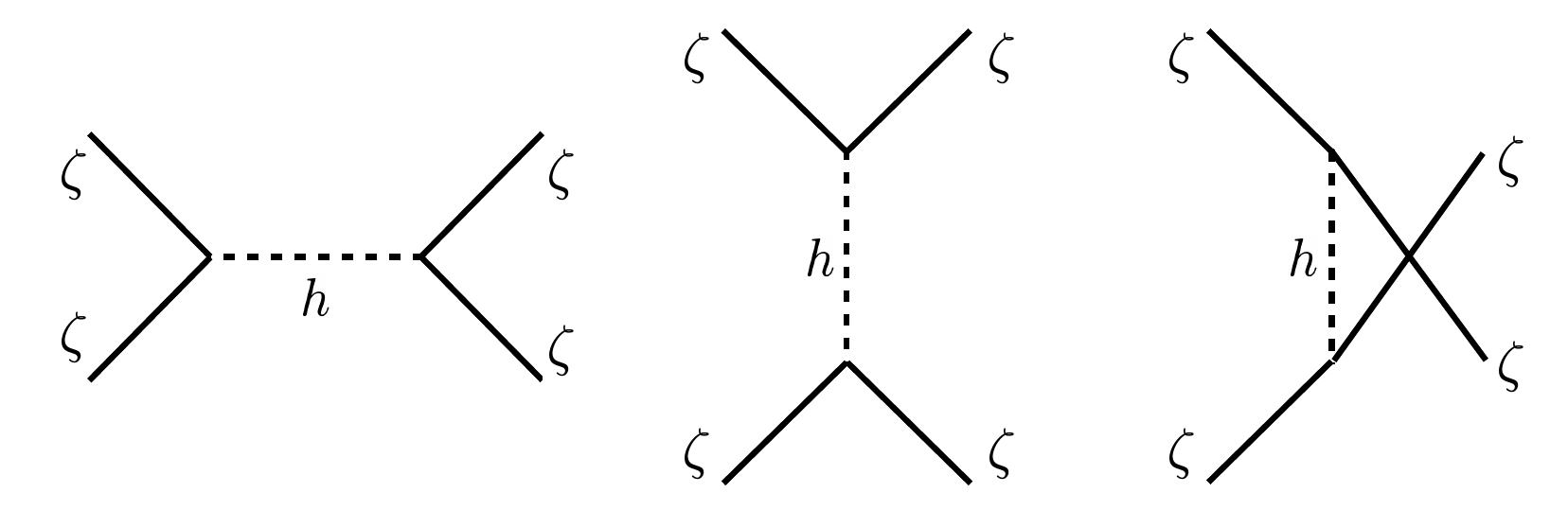}
    \caption{Tree-level diagrams with $\zeta$-legs and $h$-propagators:
      case $\textbf{(b)}$.}
    \label{fig:h-prop}
\end{figure}

{\bf (b)}. Strong enhancement would appear to occur also for diagrams with
scalar external legs and tensor exchange,
  Fig.~\ref{fig:h-prop}.
 However, as we pointed out,
the maximum number of spatial derivatives in each
$\zeta \zeta h$
vertex is 2 (rather than 4). Therefore, the  enhancement factor is
\be
u_S^{-3/2} \cdot u_S^{-3/2}\cdot 1 \cdot u_S^{-2} \cdot  u_S^{-2} = u_S^{-7} \; .
\label{jun11-22-1}
\ee
So, tensor exchange gives subdominant contribution.

\begin{figure}
    \centering
    \includegraphics[width=0.5\textwidth]{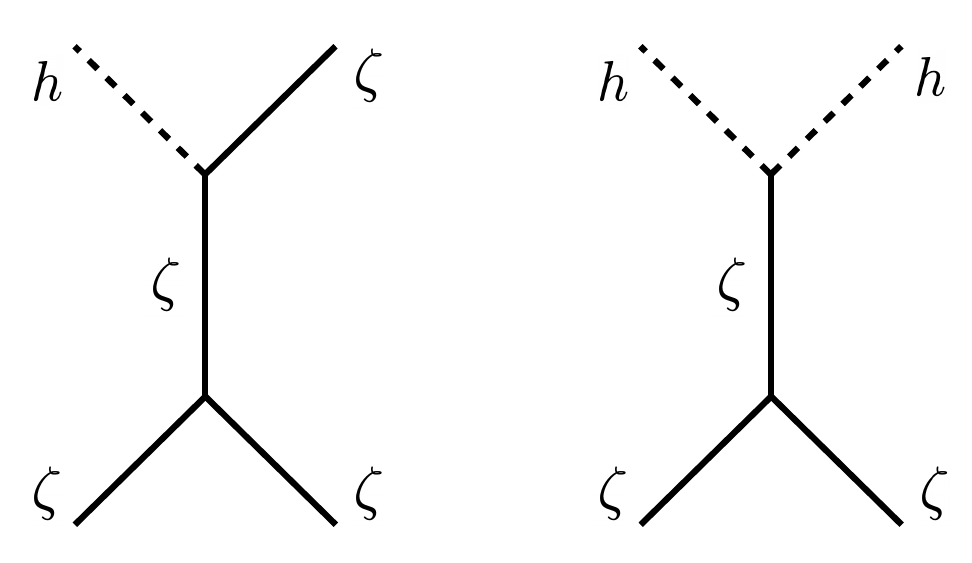}
    \caption{Tree-level t-channel diagrams:
      cases $\textbf{(c)}$ (left) and $\textbf{(d)}$
        (right).}
    \label{fig:c_d}
\end{figure}

{\bf (c)}. Process \eqref{jun10-22-2c} with $t$-channel $\zeta$-exchange,
Fig.~\ref{fig:c_d}, left diagram.
The incoming spatial momenta are of order $E$, while outgoing
ones are of order $E/u_S$. So, the spatial
momentum transfer is of order $E/u_S$ and the mechanism (iii)
does not work. The enhancement factor is
\be
u_S^{-1/2}\cdot u_S^{-3/2} \cdot 1 \cdot u_S^{-2} \cdot u_S^{-4}
=  u_S^{-8} \; ,
\label{jun11-22-2}
\ee
so this process is also subdominant.

{\bf (d)}. To illustrate the mechanism (iii), we consider
the process \eqref{jun10-22-2b} with $t$-channel
$\zeta$-exchange, 
Fig.~\ref{fig:c_d}, right diagram. Spatial momenta
of incoming and outgoing particles are of order $p \sim E$,
the spatial momentum transfer is also of order $E$,
so the mechanism (ii) does not work. We find the enhancement factor
\be
u_S^{-1/2} \cdot u_S^{-1/2} \cdot u_S^{-2} \cdot 1 \cdot 1 = u_S^{-3}\;,
\label{jun11-22-3}
\ee
which is again weak.

Other diagrams can be studied in a similar way, and all of them are
suppressed as compared to purely scalar process {\bf (a)}.
The general argument is straightforward. By replacing one or more external
scalar legs in a purely scalar diagram of
the case {\bf (a)} by  tensor leg(s),
one loses at least
an enhancement factor $u_S^{-1}$ from (i) and another factor $u_S^{-2}$
from (ii). One could in principle gain a factor $u_S^{-2}$ due to
(iii) (we have seen in our example {\bf (c)} that this is actually
not  the case if one replaces just one scalar leg),
but in any case, the overall suppression of a new
diagram is at least $u_S$ as compared to the original, purely scalar
one\footnote{In reality the suppression is always even stronger, cf.
  \eqref{jun11-22-1},  \eqref{jun11-22-2},  \eqref{jun11-22-3}.
  Moreover, 
  in mixed sectors some
couplings 
  in the cubic  action are  themselves suppressed by positive powers of
    $u_S$, leading to even 
  stronger suppression of the contributions to the amplitudes. }.
Replacing the scalar internal line by the tensor line does not improve
the situation, and in the particular case {\bf (b)} produces suppressed
result.

We conclude that
the latest time of the onset of the classical theory $t_{cl}$
is associated with the scalar sector of the theory.
This means, in particular, that the search for the largest eigenvalue
of the rescaled PWA matrix $\tilde{a}$ (Sec.~\ref{sec:speeds-unitarity})
is unnecessary. The
relevant terms in the cubic scalar action are those with four
{\it spatial} derivatives. 
Modulo numerical factors, the time $t_{cl}$
is indeed determined from \eqref{feb3-22-1},
\be
|t_{cl}|^{2\mu - 2}  \sim g u_S^{11} \; ,\nonumber
\ee
whereas for $\mu=1$ the condition is $ g u_S^{11} >1$.
To refine these estimates, we perform the calculation of the dominant
partial
wave amplitudes in the scalar sector.

\subsubsection{Strong coupling scale from tree-level unitarity}


We are now ready to perform the calculation of the strong
coupling energy scale, as implied by the tree-level unitarity bound.
The (off-shell)
cubic
action with 4 spatial derivatives
in the scalar sector is given by \eqref{jun11-22-22a}.
We continue
to use the approach of Sec.~\ref{sec:hierarchy},
set $a=1$ as before
and work with
the field $\tilde{\zeta}=   (2\mathcal{G}_S)^{-1/2} \zeta$, which has
canonical time-derivative term and gradient term with sound speed $u_S$. 
Then the cubic action
reads
 \begin{align*}
        \mathcal{S}^{(3)}_{\zeta\zeta\zeta} 
        =  \int d {t}~d^3 {x} 
        \tilde{\Lambda}_\zeta
               \partial^2 \tilde{\zeta} (\partial \tilde{\zeta})^2 \; ,
   \end{align*}
   where
   \be
   \tilde{\Lambda}_\zeta =
   (2\mathcal{G}_S)^{-3/2} \Lambda_\zeta =
   \frac{\lambda_\zeta |t|^\mu}{g_S^{3/2} g^{1/2}} \cdot |t|^2 \; .\nonumber
      \ee
      It is now straightforward to calculate the $2\to 2$
      matrix element $M (\cos \theta, E)$
      as function of scattering angle $\theta$ and center-of-mass energy $E$.
      We get
\begin{equation*}
    M = \frac{E^6}{4 u_S^8}(1-\cos^2\theta)  \tilde{\Lambda}_\zeta^2\;,
\end{equation*}
(the origin of the dependence on $u_S$ is the mechanism (ii) in
Sec.~\ref{sec:hierarchy}).
We now write the rescaled PWA amplitude
\be
\tilde{a}^{(l)} = \frac{1}{2u_S^3} \cdot
\frac{1}{32\pi}\int ~d(\text{cos}\theta)~P_l(\text{cos}\theta)\, M \; ,\nonumber
\ee
where $P_l$ are Legendre polynomials,
the factor $(2u_S^3)^{-1}$ comes from the redefinition
\eqref{jun9-22-5} (i.e.,
it  is due to the mechanism (i) in Sec.~\ref{sec:hierarchy}),
and obtain
\begin{subequations}
\begin{align*}
  \tilde{a}^{(0)} &= \frac{\tilde{\Lambda}_\zeta^2  E^6}{192\pi u_S^{11}}\;,
  \\
  \tilde{a}^{(2)} &= -\frac{  \tilde{\Lambda}_\zeta^2  E^6}{960 \pi u_S^{11}}\;.
\end{align*}
\end{subequations}
The
lowest
bound on the energy of strong coupling comes from
the $s$-wave amplitude. It saturates the unitarity bound
$|\tilde{a}^{(0)}| \leq 1/2$ at energy
\be
E_{strong} (t) =  \left(\frac{96\pi u_S^{11}}{\tilde{\Lambda}^2}\right)^{1/6}
= \frac{(96\pi)^{1/6}}{|t|} \left( \frac{g_S^3}{\lambda_\zeta^2}
\frac{g u_S^{11}}{|t|^{2\mu -2}} \right)^{1/6}.\nonumber
\ee
This refines the estimate \eqref{jun11-22-50}. Proceeding as in
Sec.~\ref{sec:srong-preliminaries}, we calculate the ratio of quantum
and classical energy scales at the time when the mode $k_* \approx k_{min}$
exits the effective horizon:
\be
\frac{E_{strong} (t_f (k_{*}))}{E_{class} (t_f (k_{*}))}
\equiv
\frac{E_{strong} (k_{*})}{E_{class} (k_{*})} 
= E_{strong} (t_f (k_{*})) \cdot |t_f (k_*)| =
C\cdot \left(\frac{u_S^8}{{\cal A}_\zeta}\right)^{1/6},
\label{jul4-22-1}
\ee
where
\be
\label{C}
C =  \frac{96^{1/6} g_S^{1/3} }{|\lambda_\zeta|^{1/3}}\left(\frac{2^{2\frac{\mu-1}{1
-\chi}}(1-\chi)^{2\frac{\mu
-\chi}{1-\chi}}(2\mu-3\chi)^{2\frac{1-\mu}{1-\chi}}}{\Gamma^2
(1-\nu)\text{sin}^2(\nu\pi)}\right)^{1/6}.
\ee
This is the desired result for general models from the class
\eqref{A_old}.
In the case $\mu=1$, $n_S=1$ (and hence $\nu = 3/2$)
we have
\be
C =  \frac{96^{1/6} g_S^{1/3} }{|\lambda_\zeta|^{1/3}} 
\left(\frac{(1-\chi)^2}{4\pi} \right)^{1/6} \; .
\nonumber
\ee
The result  \eqref{jul4-22-1} depends 
in a fairly complicated way,
 through the parameters $\chi$,
$g_S$ and $\lambda_\zeta$,
on both  the form of the Lagrangian functions ($a_2(N)$ and $a_3(N)$ in
\eqref{A_old}) and the solution to the equations of motion,
eqs.~\eqref{jan24-22-10}.
To get an idea of how restrictive the condition for the absence of strong
coupling at the horizon
exit 
is, we now turn to
concrete examples where all above points are seen explicitly.


\section{Examples}
\label{sec:explicit-bounce}

\subsection{$\mu > 1$, $n_S<1$.}
\label{sec:mu_less_1}

In this Section we consider a particular model of the type
\eqref{A_old} with the simple forms of the
Lagrangian functions:
\begin{subequations}
  \label{jul18-22-1}
\begin{align}
\label{a_2}
    a_2(N) &= c_2 + \frac{d_2}{N}\; ,\\
    \label{a_3}
    a_3(N) &= c_3+   \frac{d_3}{N} \; ,
\end{align}
\end{subequations}
where $c_2$, $c_3$, $d_2$, $d_3$, are dimensionless constants.
Making use of eqs.~\eqref{feb5-22-1}, we obtain
\begin{subequations}
  \begin{align*}
    f_S &= -2\left(\frac{4\mu -2 + d_3}{2\chi  + d_3}\right)\; ,
\\
    g_S &= \frac{6 d_3^2}{(2\chi + d_3)^2} \; .
\end{align*}
\end{subequations}
In accordance with
our discussion in Sec.~\ref{subsec:perturbations}, one
finds that the only way to obtain small
tensor-to-scalar ratio \eqref{feb1-22-2} is to ensure that $f_S \ll 1$
and $g_S \sim 1$, so that $u_S^2 = f_S/g_S \ll 1$.
We begin with the case $\mu > 1$, which corresponds to
  $n_S<1$ in accordance with the $\Lambda$CDM value \eqref{jul17-22-1}, and
  ensure the small value of $u_S$ by
imposing {\it a fine tuning relation}
\be
d_3 = -2 \; .\nonumber
\ee
This
  choice appears rather remarkable, but we do not know whether it
may be a consequence of some symmetry or dynamical principle. Anyway,
%
with this choice,
we have
\begin{subequations}
  \begin{align*}
    f_S &= \frac{4(\mu -1)}{1-\chi} = 2(1 - n_S)\; ,
\\
    g_S &= \frac{6 }{(1 -\chi)^2} \; ,
\end{align*}
\end{subequations}
where we recall  eq.~\eqref{general_n_s}.
Interestingly, small tensor-to-scalar ratio $r \sim f_S^\nu/g_S^{\nu-1}$
and small scalar tilt $(1-n_S)$ 
are now governed
by one and the same small parameter $(\mu -1)$.

Proceeding
with $d_3 = -2$, we find that
equations for the background have relatively simple form.
{These are algebraic equations:
\begin{subequations}
    \begin{align*}
  3\chi^2 - 6\chi + c_2 N^2 &=0\;,
  \\
  3\chi^2 +2 (2\mu + 1) (1 - \chi) - \kappa N + c_2 N^2 &=0\;,
    \end{align*}
    \end{subequations}
where
  {
\be
\label{kappa}
\kappa = c_3 (1+2\mu) - d_2 \; .
\ee
  The relevant solution to these equations is (the second solution
  does not yield stable bounce)
\begin{subequations}
\label{chi_N}
  \begin{align}
    \chi &= \frac{3 + 8\rho (\mu-1)(2\mu+1) - \sqrt{9 -
        12\rho (2\mu+1)(5-2\mu)}}{3 + 16\rho (\mu - 1)^2}\; ,
    \\
    N &= \frac{2}{\kappa}\left[1 + 2\mu - 2(\mu -1)\chi\right] \; ,
  \end{align}
\end{subequations}
where
 \be
 \label{rho}
    \rho = \frac{c_2}{\kappa^2}\; .
    \ee
  While the expression for $N$ is not of particular physical significance
  (the only requirement is that $N>0$),
  the value of $\chi$ is an important characteristic of the solution.
  Note that while $N$ depends on $c_2$ and $\kappa$ separately, 
  the parameter $\chi$  depends  (for given
  $\mu$)
on one combination $\rho$
  out of the
  three Lagrangian parameters remaining after setting $d_3=-2$. We will
  see in what follows that $\mu$ and $\rho$ (or, equivalently, $\chi$)
  are
  the only parameters relevant
  also for  the strong coupling issue.
    
    For small and positive $(\mu -1)$, the allowed range of
    parameters is
    \be
    \label{cond_1}
    \kappa >0 \; , \;\;\;\;
      0< \rho \lesssim \frac{2}{27} \; .
  \ee
   These relations ensure that $N>0$ and,
     importantly,
     $2\mu-3\chi>0$, see
  \eqref{jul5-22-100}.
      
    
   We are now equipped with the explicit formulas to see what
   range of the
   tensor-to-scalar ratio is consistent with our weak
   coupling calculations. We obtain from~\eqref{jun11-22-15},
   \eqref{jun12-22-10}
   \begin{subequations}
\label{jul18-22-10}
     \begin{align}
       \theta &= 1-\chi \; ,
\\
\lambda_\zeta &= \frac{1}{4(1-\chi)^2} \; ,
     \end{align}
   \end{subequations}
   while the parameter $\nu$ is still given by \eqref{feb4-22-1}.
   Besides the dependence on $\mu$,
   these parameters again depend on $\rho$ only.

   We express the parameter $\mu$  through $n_S$ and $\chi$ using
   \eqref{general_n_s}. Then we are left with the only free
   parameter $\chi$ (or, equivalently, $\rho$). Our final formulas
   are
   obtained from \eqref{feb1-22-2} and \eqref{jul4-22-1}:
   \begin{align*}
     r &= 48 (1-\chi)^{2(\nu-1)} \left( \frac{1-n_S}{3} \right)^\nu,
     \\
\frac{E_{strong} (k_{*})}{E_{class} (k_{*})}
&= E_{strong} (t_f (k_{*}) \cdot |t_f (k_*)| = \tilde{C}\cdot
\left( \frac{r^{4/\nu}}{{\cal A}_\zeta}\right)^{1/6},
     \end{align*} 
   where, as before, $\nu = 2- n_S/2$ and
  \begin{equation*}
      \tilde{C} = \frac{C}{(8g_S)^{2/3\nu}}\;,
  \end{equation*}
  where $C$ is given by \eqref{C},   so that
   \begin{align*}
     \tilde{C} 
     &= 2^{\frac{12-11n_S}{24-6n_S}}3^{\frac{4-3n_S}{24-6n_S}} (1-\chi)^{\frac{12-n_S}{3(4
     -n_S)}}\left(\frac{(2-2\chi)^{1-n_S}\Big(2+(1-n_S)-\chi\big(3
   +(1-n_S)\big)\Big)^{-(1-n_S)}}{\Gamma^2(\frac{n_S}{2}-1)
   \text{sin}^2\left[(2-\frac{n_S}{2})\pi\right]}\right)^{1/6}
     \nonumber\\
     &\approx \frac{3^{1/18}(1-\chi)^{11/9}}{2^{5/18}\pi^{1/6}}
        = 0.7 \cdot (1-\chi)^{11/9}\;,
   \end{align*}
   where we set $n_S = 1$ in the
     last two expressions. 
   In Fig.~\ref{fig:r} we  show $r$-ratio
 and the ratio  $E_{strong}(k_*)/E_{cl}(k_*)$ as 
 functions of $\chi$ for the
   $\Lambda$CDM
 central value
 $n_S = 0.9649$ suggested by observations. The main message is that
 the value of $r$ is bounded from below in our model,
$r >0.018$ for
  $n_S = 0.9649$,
 even for very generous unitarity bound $E_{strong}(k_*)/E_{cl}(k_*)>1$.
 Note that the parameters should obey $(2\mu - 3\chi)>0$ which translates to
 \be
  \label{cond_2}
 \chi < \frac{3-n_S}{4- n_S} \approx \frac{2}{3} \; .
 \ee
 This bound is also shown in Fig.~\ref{fig:r}.
 \begin{figure}[H]
\centering
\includegraphics[width=0.8\textwidth]{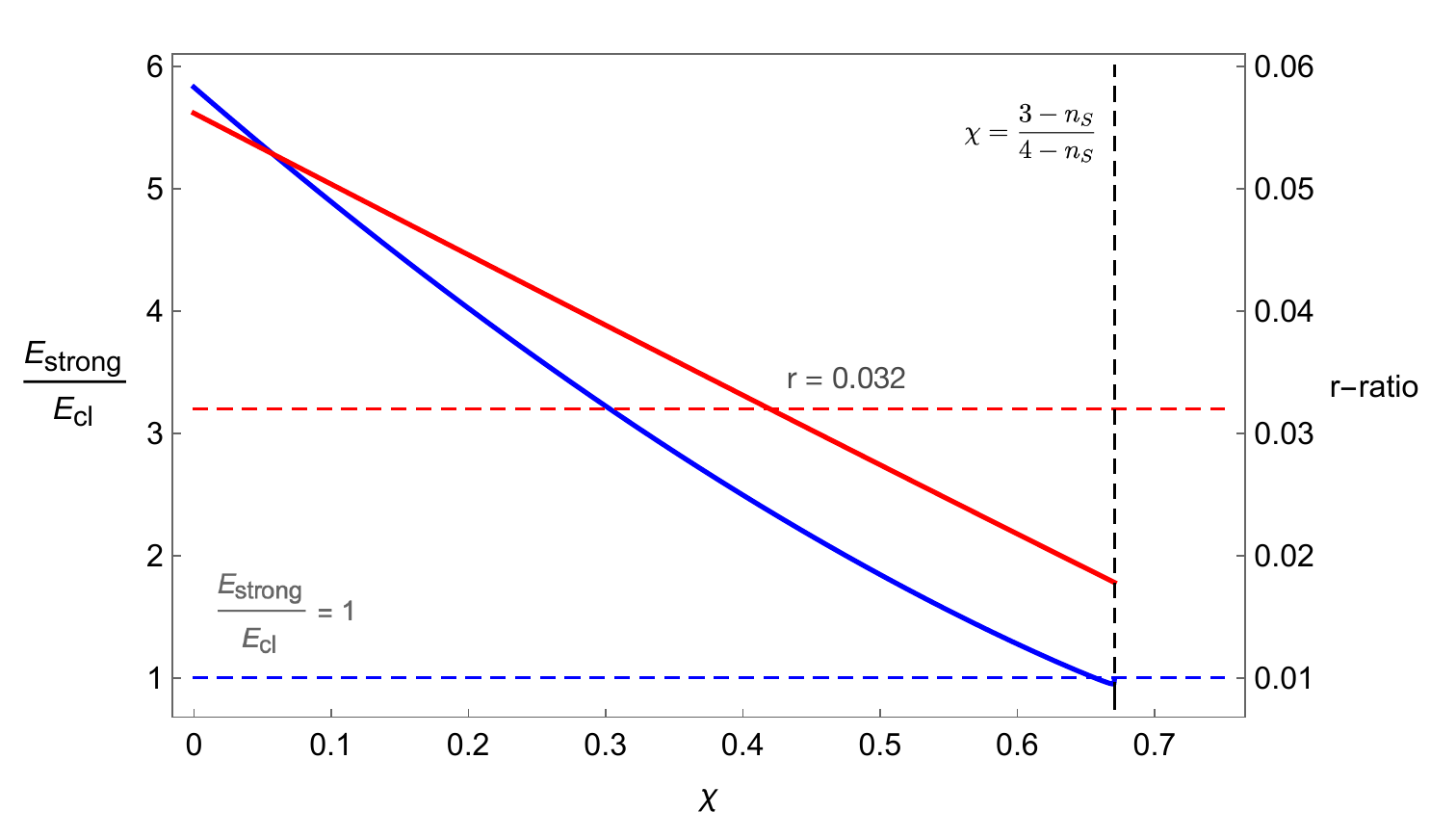}

\caption{\label{fig:r} The  ratio $E_{strong}(k_*)/E_{cl}(k_*)$  (blue line)
  and $r$-ratio (red line)
as functions of $\chi$
for the central value $n_S = 0.9649$. The allowed range $r< 0.032$ and the
unitarity bound $E_{strong}(k_*)/E_{cl}(k_*) > 1$ restrict the
parameter space to the
    right lower part.
  }
\end{figure}

   In  Fig.~\ref{fig:region} we show
     what happens if the scalar tilt $n_S$ is varied within the
     observationally allowed range. A point to note is that in the
     entire allowed range, the parameter $r$ is fairly large,
     $r > 0.015$,
     while the strong coupling scale is always close to
     the classical energy scale, $E_{strong} \lesssim 3 E_{cl}$.
     We conclude that our simple model is on the verge of being ruled out.
\begin{figure}[H]
\centering
\includegraphics[width=0.8\textwidth]{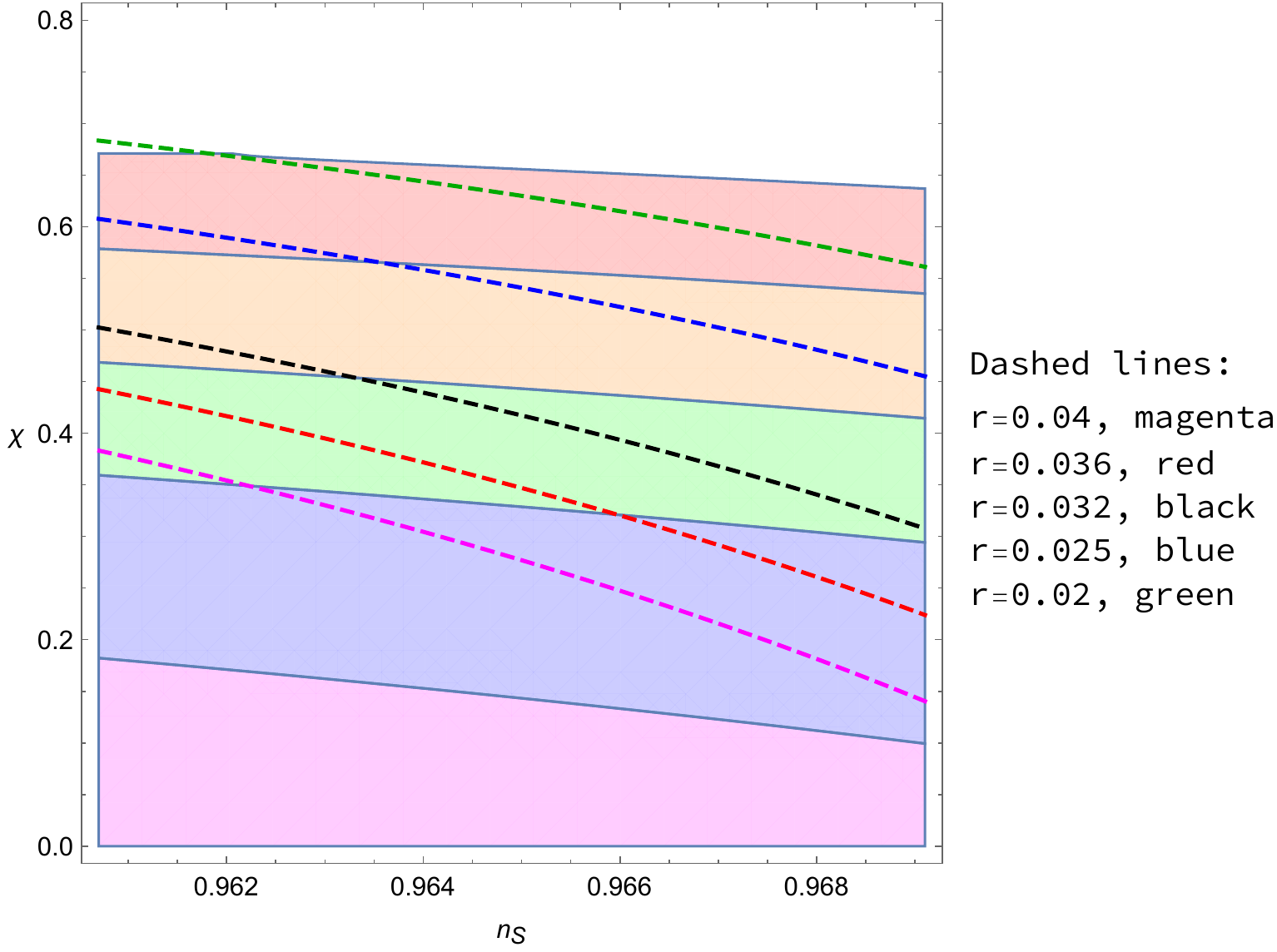}
\caption{Space of parameters $n_S$ and $\chi$. 
  Colored strips correspond to different
    ratios of strong coupling scale to classical scale:
    $1< E_{strong}(k_*)/E_{cl}(k_*)<1.5$ (red),
$1.5< E_{strong}(k_*)/E_{cl}(k_*)   <2.2$ (orange),  
$2.2< E_{strong}(k_*)/E_{cl}(k_*)  <3$ (green), 
    $3< E_{strong}(k_*)/E_{cl}(k_*) <4.5$ (blue),
    $4.5 < E_{strong}(k_*)/E_{cl}(k_*) $ (magenta).
    Dashed lines show 
   the tensor-to-scalar ratio:
    $r = 0.02$ (green),  $r = 0.025$ (blue),
    $r = 0.032$ (black),  $r = 0.036$ (red), and $r = 0.04$ (magenta).}
\label{fig:region}
\end{figure} 
Choosing appropriate values of ($\chi$, $n_S$) one immediately obtains concrete form of scale factor \eqref{jan31-22-1} and Hubble parameter $H = \chi/t$ in Jordan frame for the contraction stage. For example, $a(t)/d$ and $H$ for $\chi = 0.4$, $n_S = 0.967$, (and i.e. for $\mu = 1.0099$) are shown in Fig.~\ref{fig:aH_J_mu_ne_1}.
\begin{figure}[h!]
\centering 
\includegraphics[width=7.5cm]{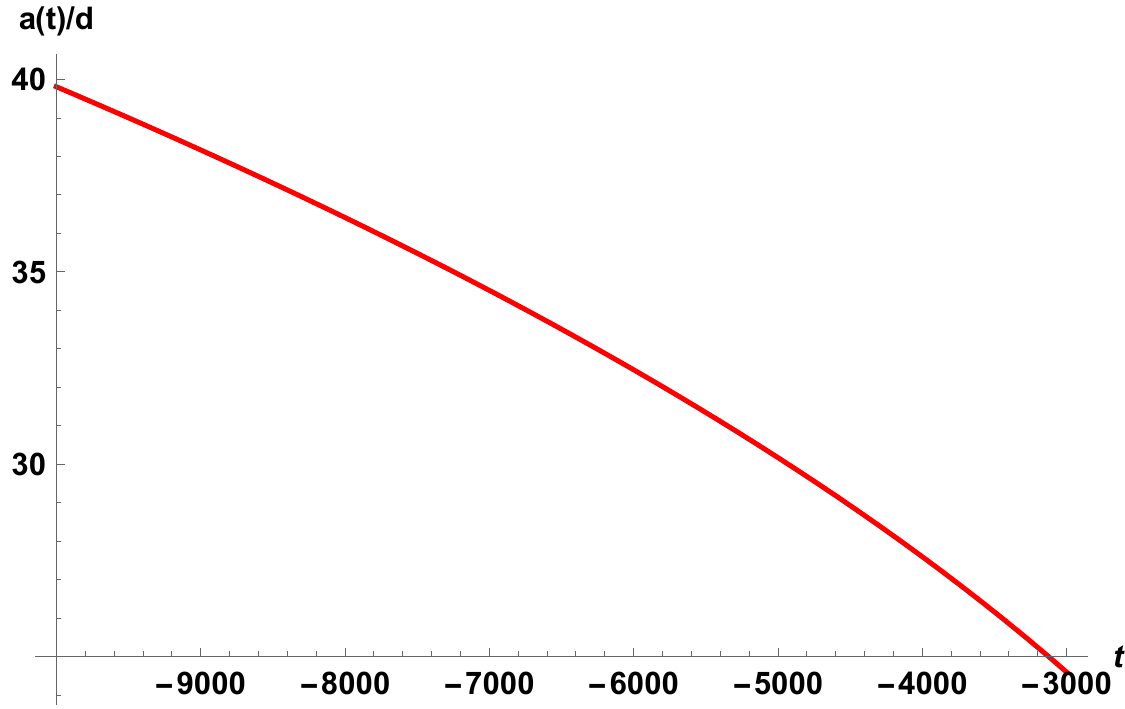}
\includegraphics[width=7.5cm]{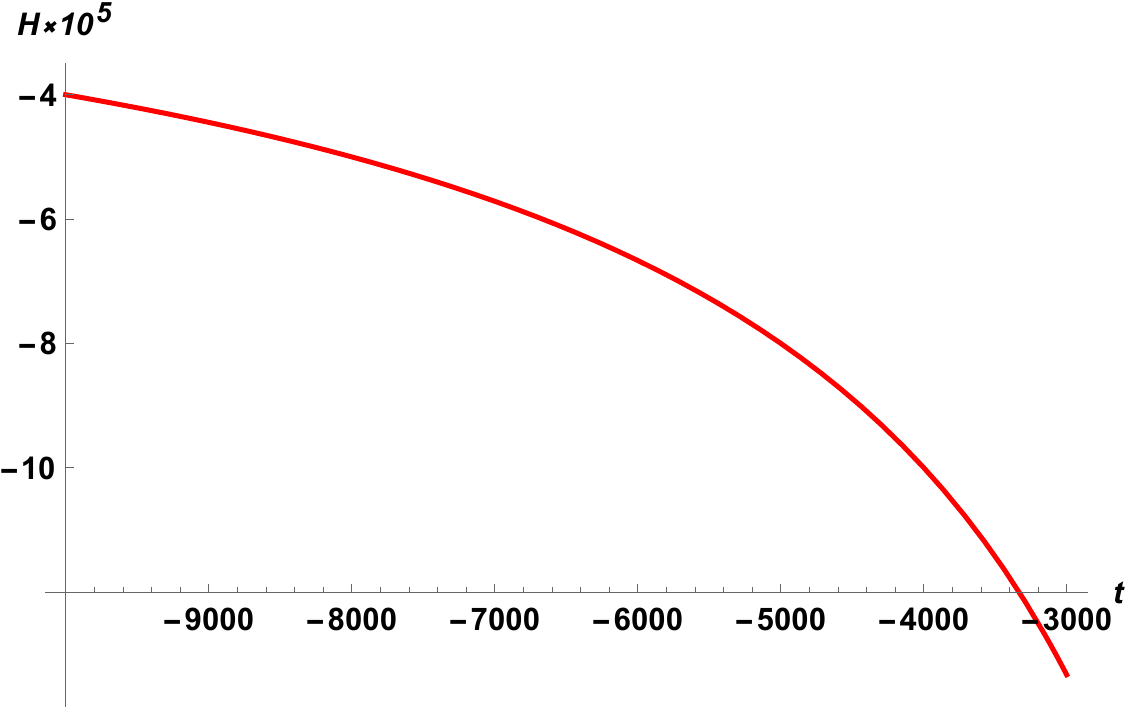}
 \caption{The scale factor $a(t)$ divided by $d$ (left panel) and the Hubble parameter $H(t)$ (right panel) in Jordan frame in the model with $\chi = 0.4$, $n_S = 0.967$ (and $\mu = 1.0099$).}
 \label{fig:aH_J_mu_ne_1}
\end{figure}
It is also interesting to find corresponding scale factor and Hubble parameter in Einstein frame. The scale factor $a_E(t_E)$ ($t_E$ is a cosmic time in Einstein frame) is given by \eqref{jun23-22-1}, while Hubble parameter in Einstein frame is
\begin{equation*}
    H_E \equiv \frac{1}{a_E}\frac{d a_E}{d t_E} = \left(\frac{\mu-\chi}{\mu-1}\right)\frac{1}{t_E}.
\end{equation*}
These scale factor and Hubble parameter with $\chi=0.4$ and $n_S = 0.967$ ($\mu = 1.0099$) in Einstein frame are shown in Fig.~\ref{fig:aH_E_mu_ne_1}.
\begin{figure}[h!]
\centering 
\includegraphics[width=7.5cm]{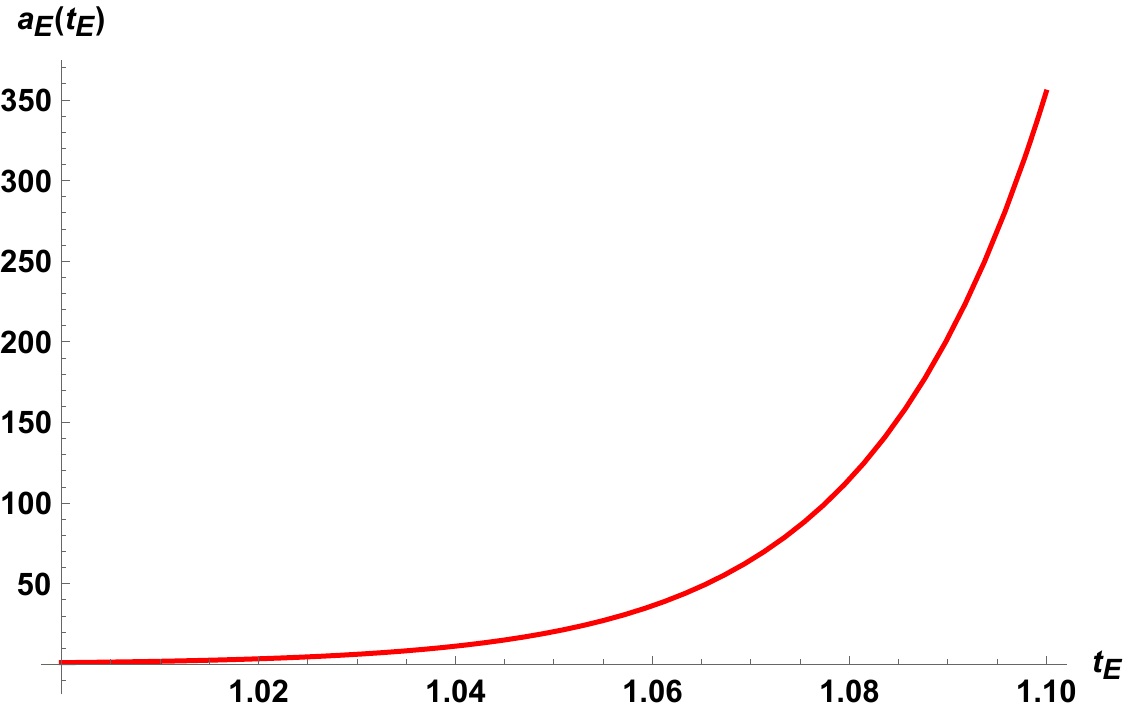}
\includegraphics[width=7.5cm]{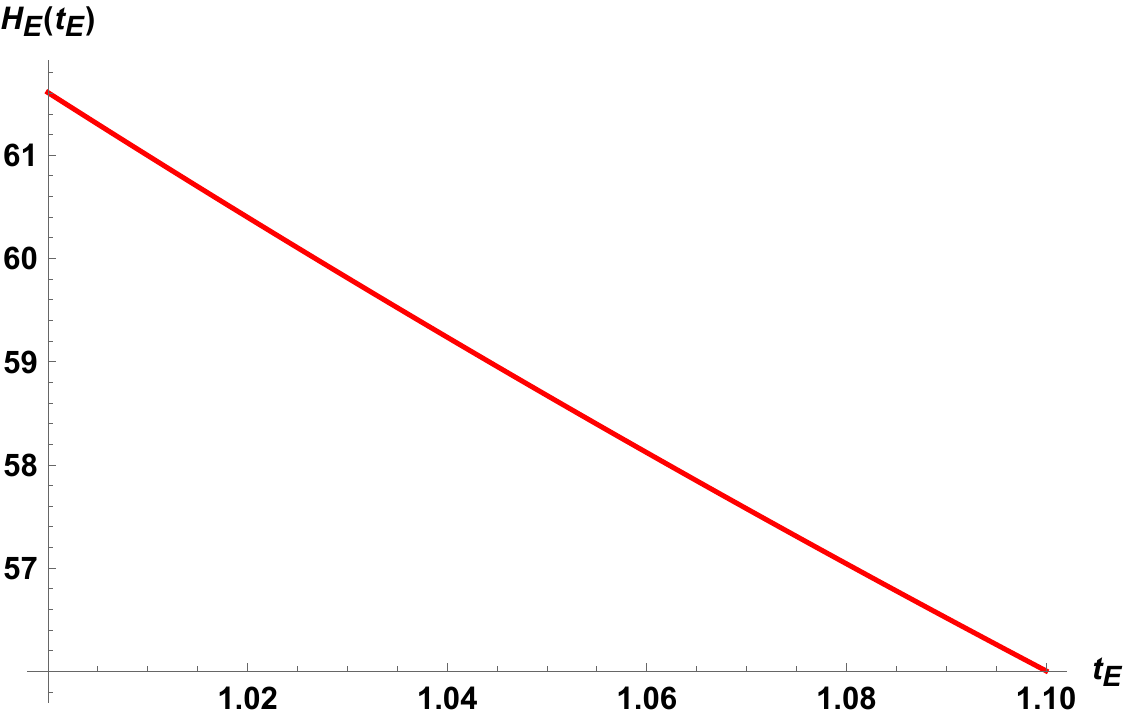}
 \caption{The scale factor $a_E(t_E)$  (left panel) and the Hubble parameter $H_E(t_E)$ (right panel) in Einstein frame in the model with $\chi = 0.4$, $n_S = 0.967$ (and $\mu = 1.0099$).}
 \label{fig:aH_E_mu_ne_1}
\end{figure}

Finally, in Appendix F we present an explicit numerical example of  stable and subluminal evolution where the contraction from this section terminates and expansion begins (bounce), then the Universe turns  into the kination epoch at future times. In this model, the Einstein-Hilbert action for gravity is restored during the kination epoch and thus all background quantities such as the energy
scales, the Hubble parameter can be calculated in a usual way.  Similar procedure was done in Ref.~\cite{Ageeva:2021yik}, where different models of bouncing Universe and genesis turn into general relativity with a conventional massless scalar field that drives the expansion in the future, at large positive times. The kination epoch is assumed to end up with reheating through, for example, one of the mechanisms of Refs.~\cite{Armendariz-Picon:1999hyi,BazrafshanMoghaddam:2016tdk}.  We also prove for our concrete numerical example (in the same Appendix F), that the proposed evolution is physically relevant, since around 60 e-folds of contraction can be accommodated so the wavelength of primordial fluctuations stretches  $\sim$60 Hubble scales at the end of the contracting phase and before the Universe bounces.

  \subsection{$\mu=1$, $n_S=1$}
\label{sec:mu1}
 We now consider the case $\mu=1$, $n_S=1$ consistent with
  the  early dark energy idea~\cite{Ye:2021nej,Jiang:2022uyg}.
  Our model is still defined by the functions \eqref{jul18-22-1},
  now with
  $\mu=1$.
  Unlike in Sec.~\ref{sec:mu_less_1},
  we ensure that $f_S\ll 1$, and hence the scalar sound speed is small,
  by choosing
  \begin{equation*}
    d_3 = -2 + \epsilon \; , \;\;\;\;\;\; \epsilon \ll 1 \; .
  \end{equation*}
  Then
\begin{align*}
    f_S &= \frac{2 \epsilon}{2-2\chi -\epsilon}\;, \\
    g_S &= \frac{6(2-\epsilon)^2}{(2-2\chi - \epsilon)^2} \; ,
\end{align*}
while $\theta$ and $\lambda_\zeta$ are still given by \eqref{jul18-22-10}. 
Note that $\epsilon>0$, since we
  require $f_S>0$. 

Background equations of motion are 
again algebraic, and their solution is
\begin{align*}
    \chi &=  \frac{(2-\epsilon)\left(1+6\epsilon \rho -\sqrt{1-12(1-\epsilon)\rho}\right)}{2(1+3\epsilon^2\rho)}\;,\\
    N &= \frac{3(2-\epsilon)\left(2-\epsilon(1-\sqrt{1-12(1-\epsilon)\rho})\right)}{2\kappa(1 + 3\epsilon^2\rho)}\;,
\end{align*}
with  $\kappa = 3c_3 - d_2$, $\rho = c_2 /\kappa^2$.
For given $\epsilon$, we can again trade the parameter $\rho$ for $\chi$,
hence the two relevant parameters are now $\chi$ and $\epsilon$.
%
Having all formulas above, we are ready to understand what range of the 
$r$-ratio is consistent with the weak coupling regime.
We recall \eqref{feb1-22-2} and \eqref{jul4-22-1}, set $\nu =3/2$ and find
\begin{equation*}
    r = \frac{16 \epsilon^{3/2}}{[3(2-2\chi -\epsilon)]^{1/2}(2-\epsilon)}\;,
\end{equation*}
\begin{equation*}
    \frac{E_{strong} }{E_{class} }
 = \tilde{B}\cdot
\left( \frac{r^{8/3}}{{\cal A}_\zeta}\right)^{1/6}\;,
\end{equation*}
where the coefficient $\tilde{B}$ is
\begin{align*}
    \tilde{B} = 6^{1/18} (1-\chi) \left(\frac{ \left[\frac{2-2\chi - \epsilon}{
    2-\epsilon}\right]^{4/3}}{4\pi}\right)^{1/6} = 0.7  \cdot (1-\chi) \left( 
   \frac{2-2\chi - \epsilon}{2-\epsilon}\right)^{2/9}.
\end{align*}

In Fig. \ref{fig:mu1} we show  $\frac{E_{strong} }{E_{class} }$ and $r$
on the plane of the 
two parameters $\epsilon$ and $\chi$. One observes that in the model
with $\mu = 1$ it is easier 
to obtain 
small tensor-to-scalar ratio, still insisting on the
generation of the perturbations in the weak 
coupling regime. Nevertheless, the value of $r$ cannot be much smaller
than 0.01, otherwise one would face with the strong coupling problem. 
 
\begin{figure}[H]
\centering
\includegraphics[width=0.8\textwidth]{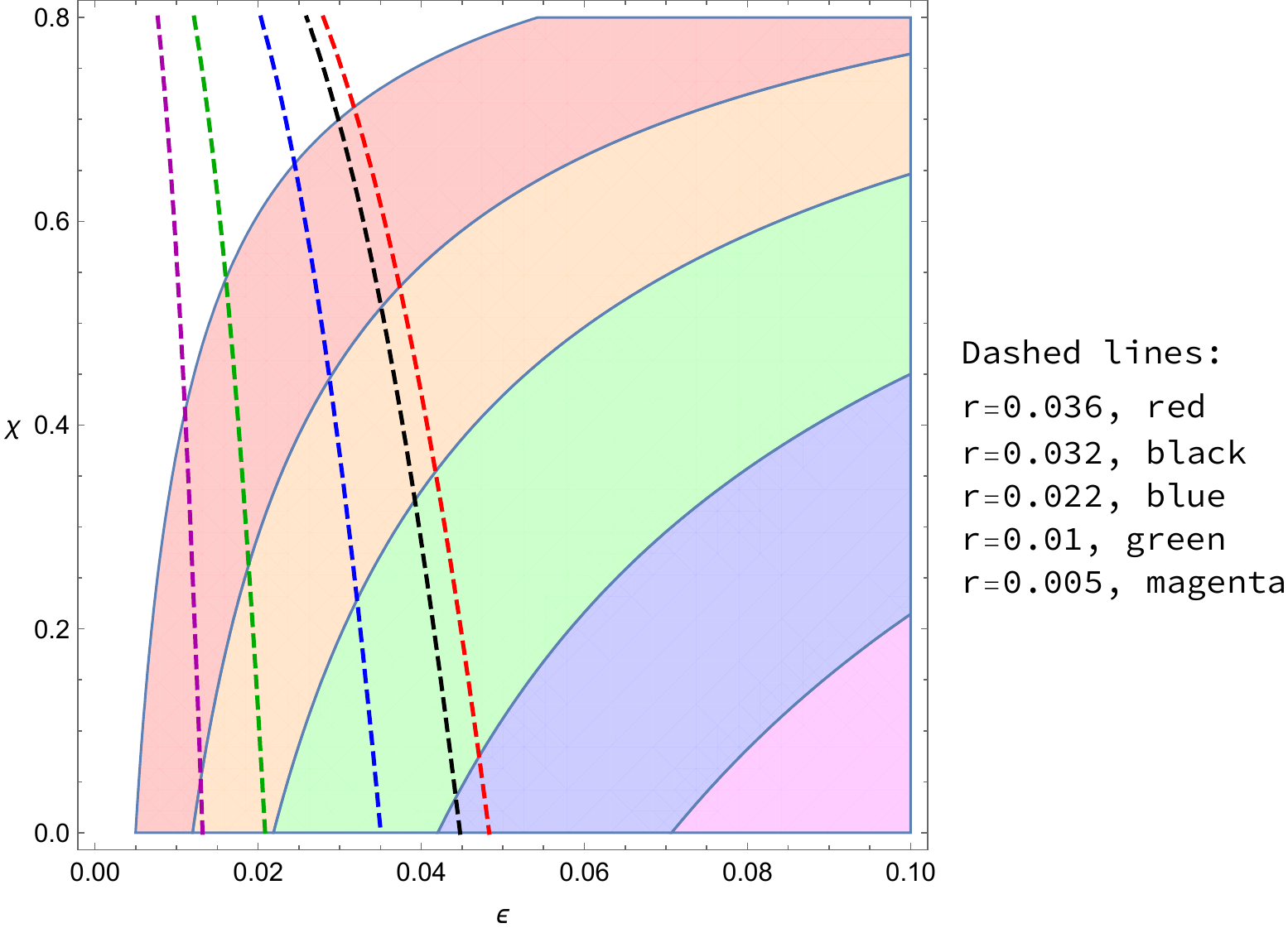}
\caption{Space of parameters $\epsilon$ and $\chi$ in the case  $\mu = 1$. 
  Colored strips correspond to different
    ratios of strong coupling scale to classical scale:
    $1< E_{strong}/E_{cl}<1.8$ (red),
$1.8< E_{strong}/E_{cl}  <2.7$ (orange),  
$2.7< E_{strong}/E_{cl}  <4.2$ (green), 
    $4.2< E_{strong}/E_{cl} <6$ (blue),
    $6 < E_{strong}/E_{cl}$ (magenta).
    Dashed lines show 
   the tensor-to-scalar ratio:
    $r = 0.005$ (magenta), $r=0.01$ (green),  $r = 0.022$ (blue),
    $r = 0.032$ (black), and  $r = 0.036$ (red).}
\label{fig:mu1}
\end{figure}

Finally, choosing appropriate value of $\chi$ in this case one can easily find the form of scale factor \eqref{jan31-22-1} and Hubble parameter $H = \chi/t$ in Jordan frame. For example, $a(t)/d$ and $H$ for $\chi = 0.1$ are shown in Fig.~\ref{fig:a_J_mu_1}.
\begin{figure}[h!]
\centering 
\includegraphics[width=7.5cm]{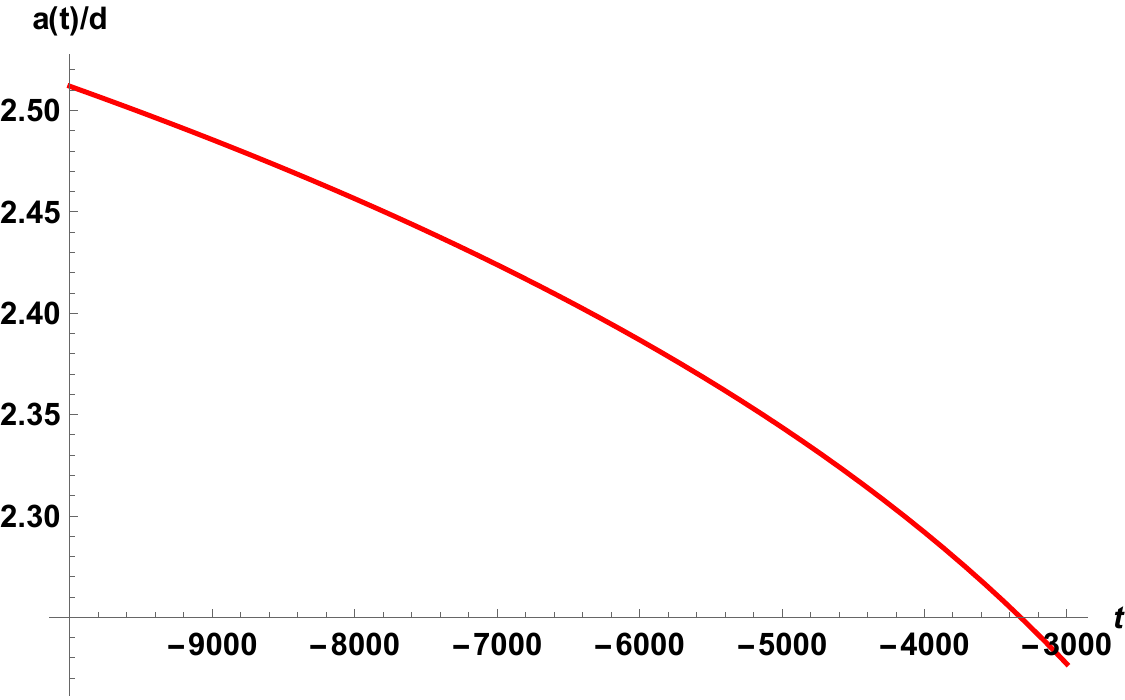}
\includegraphics[width=7.5cm]{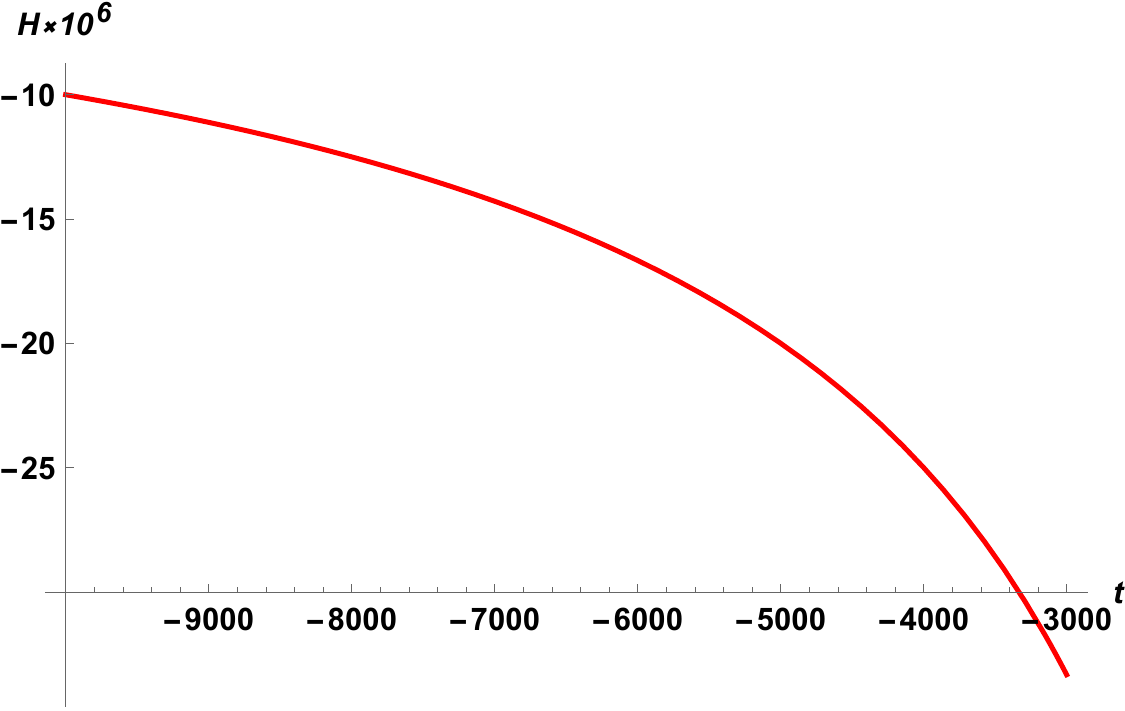}
 \caption{The scale factor $a(t)$ divided by $d$  (left panel) and the Hubble parameter $H(t)$ (right panel) in Jordan frame in the model with $\chi = 0.1$, $n_S=1$, and $\mu=1$.}
 \label{fig:a_J_mu_1}
\end{figure}

Here we give scale factor and Hubble parameter in Einstein frame as well
\begin{equation*}
    a_E  =   \frac{g_1}{(1-\chi)} d^{1/\chi} e^{(1-\chi)\frac{t_E}{g_1}},
\end{equation*}
    \begin{equation*}
    H_E = \frac{d a_E}{a_Ed t_E} = \frac{1-\chi}{g_1},
\end{equation*}
with $g_1 \equiv \frac{\sqrt{g}}{M_{Pl}}$ and $t_E$ is a cosmic time in Einstein frame.  We discuss all related calculations of these scale factor and Hubble parameter in Einstein frame in Appendix E. 
Scale factor (divided by some convenient constant) in Einstein frame with chosen $\chi=0.1$ is shown in Fig.~\ref{fig:a_E_mu_1}. 
\begin{figure}[h!]
\centering 
\includegraphics[width=8.5cm]{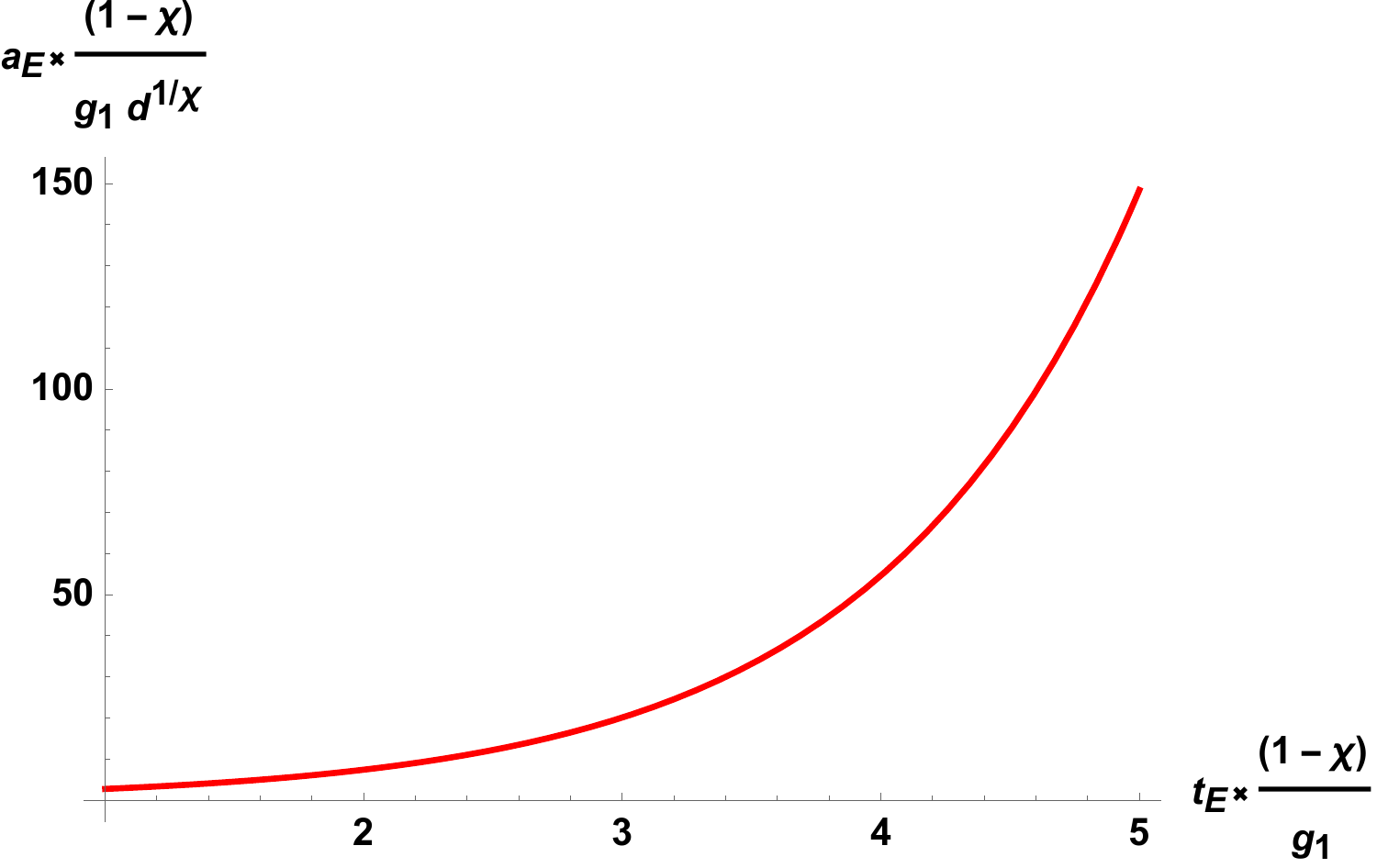}
 \caption{The scale factor $a_E(t_E)$ (multiplied by some convenient constant)  in Einstein frame in the model with $\chi = 0.1$, $n_S =1$, and $\mu=1$.}
 \label{fig:a_E_mu_1}
\end{figure}

\section{Conclusion}
\label{sec:conclude}

In this paper we have studied,  
   in the 
framework of the Horndeski theory, the contracting cosmological
stage which can afterwards pass through bounce
to a realistic cosmological expansion, as discussed in detail, e.g., in
Ref.~\cite{Ageeva:2021yik}. We have found that this stage
is capable of producing experimentally 
consistent scalar power spectrum and small enough
tensor-to-scalar ratio $r$. Small value of $(1-n_S)$ is obtained at the expense
of mild fine tuning, while small $r$ requires small scalar sound speed,
$r\sim u_S^3$. The latter property is in potential tension with
the requirement of 
   weak coupling
at
the time of the generation of scalar perturbations: small $u_S$ enhances the
scattering amplitudes and modifies the unitarity relation, which is violated
at energies dangerously close to the energy scale of the classical evolution.
Thus, very small values of $r$ are strongly disfavored in our class of models.
We have illustrated these properties in
very concrete examples within
the Horndeski theory.

While our motivation originates from the interest in constructing
complete, singularity-free cosmologies, our initial stage of the bounce
is conformally equivalent to rapidly expanding Universe, and, indeed,
the red scalar tilt is obtained in a model, conformally
equivalent\footnote{The action for our set of models has fairly simple form
  in our Jordan frame and is a lot more contrived in the inflationary
  Einstein frame. For this reason we have performed our analysis entirely
  in the Jordan frame.} to
$G$-inflation~\cite{Kobayashi:2010cm}.
So, we  expect that our observation of the
importance of the quantum strong coupling problem may be relevant to
the models of $G$-inflation, and possibly also k-inflation. We think this
line of thought is worth pursuing in the future.

Finally, it is known, that models of galileon genesis and k-inflation typically produce non-gaussian features incompatible with current observations of the CMB. It would be interesting to examine this problem for the contracting models, one of which was studied in this paper. We plan to turn to this issue in the upcoming works.

   \section*{Acknowledgemets}
   The authors are grateful to M. Shaposhnikov,
   A. Starobinsky,
   A. Westphal, and C. Wetterich
   for helpful
   discussions
and Yun-Song Piao for useful correspondence.
   This work has been supported by Russian Science Foundation
   Grant No. 19-12-00393.

\section*{ Appendix A. General expressions in the Horndeski model}
Here we give explicit formulas for a theory with action \eqref{adm_lagr}.
Equations of motion for spatially flat FLRW background read \cite{Kobayashi:2011nu}
\begin{subequations}
\label{eoms}
    \begin{eqnarray}
         &  (NA_2)_N + 3NA_{3N}H + 6N^2(N^{-1}A_4)_N H^2 = 0\;,\\
      &  A_2
        -6A_4H^2-\frac{1}{N}\frac{d}{d\hat{t}}\left( A_3+4A_4H \right) = 0 \; ,
    \end{eqnarray}
\end{subequations}
where $H = (Na)^{-1} (da/d\hat{t}$) is the Hubble parameter.
The coefficient functions in the quadratic actions for perturbations \eqref{jan23-22-5}
are given by \cite{Kobayashi:2015gga}
\begin{subequations}
  \label{eq:Ft_Gt_form}
\begin{align}
         \mathcal{ G}_T &=  -2A_4\;,\\
         \mathcal{ F}_T &= 2B_4\;,
\end{align}
\end{subequations}
 and
\begin{subequations}
    \label{eq:Fs_Gs_form}
    \begin{eqnarray}
        \mathcal{ F}_S&=&\frac{1}{a N}\frac{d}{d \hat{t}}\left(\frac{a}{\Theta}\mathcal{ G}_T^2\right)
        -\mathcal{ F}_T\;, 
        \label{eq:Fs_form}
        \\
        \mathcal{ G}_S&=&\frac{\Sigma }{\Theta^2}\mathcal{ G}_T^2+3\mathcal{ G}_T\;, \label{eq:Gs_form}
    \end{eqnarray}
    \end{subequations}
 with
\begin{subequations}
 \begin{align}
      \Sigma&=
      N A_{2N}+\frac{1}{2}N^2A_{2NN}+
      \frac{3}{2}N^2A_{3NN}H+3\big(2A_4-2NA_{4N}+N^2A_{4NN}\big)H^2\;, 
        \\
        \Theta&=2H\Big(\frac{NA_{3N}}{4H}-A_4 + NA_{4N}\Big)\;.
        \label{theta}
    \end{align}
\end{subequations}

\section*{Appendix B. Power spectra}

In this Appendix we present, for completeness, the calculation
of the power spectra of perturbations. The quadratic actions
for perturbations are given by eqs.~\eqref{jan21-24-2}, where
the scale factor is written in eq.~\eqref{jan31-22-1} and the
coefficients are given by eqs.~\eqref{jan31-22-2}, \eqref{jan31-22-3}.

We give the calculation for the scalar perturbations; tensor
perturbations are treated in the same way.
We introduce the canonically normalized field $\psi$ via
\begin{equation*}
\zeta=\frac{1}{\left(2\mathcal{G}_S a^{3}\right)^{1 / 2}} \cdot \psi\;,
\end{equation*}
so that the quadratic action is
\begin{equation*}
  \mathcal{S}_{\psi \psi}^{(2)}=\int d^{3} x d t
  \left[\frac{1}{2} \dot{\psi}^{2}+\frac{1}{2}
    \frac{\ddot{\alpha}}{\alpha} \psi^{2}-\frac{u_S^2}{2a^2} (\vec{\nabla} \psi)^{2}\right]\;,
\end{equation*}
where 
\begin{equation*}
  \alpha=\left(2\mathcal{G}_S a^{3}\right)^{1 / 2} =
  \frac{\mbox{const}}{(-t)^{\frac{2\mu - 3\chi}{2}}} \; .
\end{equation*}
Once the inequality
\eqref{jul5-22-100} is satisfied,
the second term in the integrand is negligible at early times
$t \to -\infty$, and the field $\psi$
can be treated within the WKB approximation.
The properly normalized negative-frequency
WKB solution is
\be
  \psi_{WKB} =\frac{1}{(2 \pi)^{3 / 2}} \frac{1}{\sqrt{2 \omega}} \cdot
  \mathrm{e}^{- i \int \omega dt}  
=\frac{1}{(2 \pi)^{3 / 2}} \sqrt{\frac{d}{2 u_{S} k}} (-t)^{\chi / 2}
\cdot \mathrm{e}^{ i \frac{u_{S}}{d} \frac{k}{1-\chi} (-t )^{1-\chi}},\nonumber
\ee
where
\begin{equation*}
\omega = \frac{u_S k}{a}=
\frac{u_S\cdot  k}{d (-t)^{\chi}}\;,
\end{equation*}
We now solve the complete equation \eqref{jan31-22-5}
for perturbation $\zeta$ with the early time asymptotics
$\zeta \to \zeta_{WKB} = \left(2\mathcal{G}_S a^{3}\right)^{-1 / 2}
\psi_{WKB}$ and obtain
\begin{equation*}
\zeta=\mathfrak{C} \cdot (-t)^{\delta} \cdot H_{\nu}^{(1)}\left(\beta (-t)^{1-\chi}\right)\;,
\end{equation*}
where
\begin{align*}
\delta &=\frac{1+2 \mu-3 \chi}{2}\;, \\
\beta &=\frac{u_{S} k}{d(1-\chi)}\;, \\
\nu &=\frac{\delta}{\gamma}=\frac{1+2 \mu-3 \chi}{2(1-\chi)}\;,
\end{align*}
and normalization factor $\mathfrak{C}$ is  found by matching
to the WKB solution;
modulo an irrelevant time-independent phase, we have
\begin{equation*}
  \mathfrak{C}=\frac{1}{(g g_S)^{1 / 2}}
  \frac{1}{2^{5 / 2} \pi(1-\chi)^{1 / 2}} \frac{1}{d^{3 / 2}}\;.
\end{equation*}
At late times (formally, $|t| \to 0$), this solution is
time-independent,
\begin{equation*}
  \zeta=
  (- i) \frac{\mathfrak{C}}{\sin (\nu \pi)} \frac{(1-\chi)^{\nu}}{u_{S}^{\nu} \Gamma(1
  -\nu)}\left(\frac{2 d}{k}\right)^{\nu}.
\end{equation*}
It determines the scalar power spectrum via
\begin{equation*}
\mathcal{P}_{\zeta}=4 \pi k^{3} \zeta^{2}\;.
\end{equation*}
Collecting all factors, we obtain the result quoted in
\eqref{general_ampl}.

Tensor power spectrum is obtained by replacing 
$2\mathcal{G}_S \to \mathcal{G}_T/4$
(i.e., $g_S \to 1/4$) and $u_S \to u_T=1$,
and multiplying by 2 due two two tensor polarizations.
This gives the result for $\mathcal{P}_T$ quoted in
\eqref{general_ampl}.

\section*{Appendix C. Largest terms in cubic actions}

In this Appendix we collect the expressions for cubic
actions that contain the largest number of {\it spatial}
derivatives.
As we discuss in the main text, we consistently work
  off-shell, i.e., do not use equations of motion for dynamical
  perturbations $\zeta$, $h_{ij}$ when
  evaluating the unconstrained cubic action.
  Neither we employ field redefinitions to get rid
    of the terms in the cubic action which are
    proportional to the linearized field
    equations; we only use the background equations of motion and
    perform integrations by parts. This is precisely what is done in
    Refs.~\cite{DeFelice:2011zh,Gao:2011qe,Ageeva:2020gti}.
  In this approach, no
  coefficients in the cubic action are enhanced by inverse powers of $u_S$. 
The terms with  the largest number of spatial
derivatives are readily extracted from
Ref.~\cite{Ageeva:2020gti}. We stick to the action \eqref{Hor_L} with $G_5=0$, or, 
equivalently, the action~\eqref{adm_lagr}; furthermore,
in our set of models we have
$G_4 = G_4 (\phi)$, or, equivalently,
\be
A_4 = - B_4 = -B_4 (\hat{t})\;,\nonumber
\ee
(no dependence on $N$, see eq.~\eqref{A4old}).
At a given moment of time we rescale spatial coordinates to set
$a=1$ as we do  in Sec.~\ref{sec:srong-preliminaries} and later in the main
text. We work in cosmic time with $N=1$.

We consider various cubic terms in turn.

\subsection*{C1. Triple-$\zeta$ action}

In the purely scalar sector, the maximum number of spatial derivatives
in the cubic action is 4, and the relevant terms, as given in
Ref.~\cite{Ageeva:2020gti}, are
 \begin{align*}
        \mathcal{S}^{(3)}_{\zeta\zeta\zeta} 
        &=  \int d {t}~d^3 {x} \left[
                  \Lambda_7 \dot{\zeta} \left( {\partial}^2 \zeta \right)^2 
        + \Lambda_8 \zeta \left( {\partial}^2 \zeta \right)^2 
        + \Lambda_9  {\partial}^2 \zeta \left( {\partial}_i \zeta \right)^2
        \right.
        \nonumber \\
        &~~~~~~~~~~~~~~~~~\left. + \Lambda_{10} \dot{\zeta} \left( {\partial}_i 
        {\partial}_j \zeta \right)^2 
        + \Lambda_{11} \zeta \left( {\partial}_i  {\partial}_j \zeta \right)^2
     \right] \; ,
    \end{align*}
 where, as before, $\partial^2 = \partial_i \partial_i$ is the spatial Laplacian, and
 \begin{subequations}
   \label{jun11-22-10}
 \begin{align}
            & \Lambda_8 = - \Lambda_{11} = 
        -\frac{3\mathcal{ G}_T^3}{2\Theta^2}\;, 
\label{jun11-22-21}
        \\
        & \Lambda_9 =
        -\frac{2\mathcal{ G}_T^3}{\Theta^2} \;,
 \end{align}
 \end{subequations}
 with $\Lambda_7 = - \Lambda_{10}$.
 The function $\Theta$,   entering \eqref{jun11-22-10}, is given by
\eqref{theta}, and
for the solution in Sec.~\ref{sec:solution-powerlaw}, the expression
for $\Theta$ reduces to \eqref{jun11-22-15}.

One notices
that terms with $\Lambda_7$ and $\Lambda_{10}$ cancel
out upon integration by parts 
(using  $\Lambda_7 = - \Lambda_{10}$
and neglecting the term of order
  $\dot{\Lambda}_{10} \, \partial_i\partial_j  \zeta\,  \partial_i \zeta \, \partial_j\zeta$).
Moreover,  among the remaining three monomials, only two are independent,
modulo integration by parts, since
\be
\int~d^3x~ \zeta \left( {\partial}_i  {\partial}_j \zeta \right)^2 =
\int~d^3x \left[ \zeta \left( {\partial}^2 \zeta \right)^2 
  + \frac{3}{2}
  {\partial}^2 \zeta \left( {\partial}_i \zeta \right)^2 \right].\nonumber
  \ee
  Making use of \eqref{jun11-22-21}, one finds that the
   relevant part of the
  triple-$\zeta$
  action has just one term
   \begin{align}
        \mathcal{S}^{(3)}_{\zeta\zeta\zeta} 
        =  \int d {t}~d^3 {x} 
        \Lambda_\zeta
               {\partial}^2 \zeta \left( {\partial}_i \zeta \right)^2 \; ,
\label{jun11-22-22}
   \end{align}
   where
   \be
   \Lambda_\zeta = \Lambda_9 - \frac{3}{2} \Lambda_8 =
   \frac{\mathcal{ G}_T^3}{4\Theta^2} \; .\nonumber
   \ee

\subsection*{C2. $h \zeta \zeta$, $hh \zeta$ and triple-$h$ actions}

In general Horndeski theory with $G_5 \neq 0$, and/or
$G_4 = G_4 (\phi, X)$, the cubic $h \zeta \zeta$ action 
has the following general form (see Ref.~\cite{Gao:2012ib} where explicit
expressions are given)
 \begin{align}
    \label{ssh}
    \mathcal{ S}^{(3)}_{\zeta \zeta h} &=  \int d {t}~d^3 {x}
    \left[ c_1
        h_{ij}\zeta_{,i}\zeta_{,j}
        +c_2 \dot h_{ij} \zeta_{,i}\zeta_{,j}
        +c_3 \dot h_{ij}\zeta_{,i}\psi_{,j} \right.
        \nonumber \\
        &~~~~~~~~~~~~~~~+\left. c_4 \partial^2h_{ij}\zeta_{,i}\psi_{,j}
        +c_5\partial^2 h_{ij}\zeta_{,i}\zeta_{,j}
        +c_6\partial^2 h_{ij}\psi_{,i}\psi_{,j}\right] \; ,
    \nonumber
 \end{align}
 where
 \begin{equation*}
        \psi=\partial^{-2} \partial_t \zeta \; .
 \end{equation*}
 The term with $c_5$ involves 4 spatial derivatives. However,
 in the particular case that we consider, $G_5=0$, $G_4=G_4(\phi)$, we have
 \begin{equation*}
    c_4 = c_5 = 0\;.
 \end{equation*}
 So,  the cubic $h \zeta \zeta$ action involves 2 spatial derivatives only.
 
 The general structure of the cubic $\zeta h h$ action is~\cite{Gao:2012ib}
  \begin{align*}
        \mathcal{ S}^{(3)}_{\zeta hh}  &=  \int d {t}~d^3 {x}
    \left[
        d_1\zeta\dot h_{ij}^2
        +\frac{d_2}{a^2}\zeta h_{ij,k}h_{ij,k}
        +d_3\psi_{,k}\dot h_{ij}h_{ij,k} +  d_4\dot\zeta\dot h_{ij}^2 \right.
        \nonumber
        \\
        &\left. +\frac{d_5}{a^2}\partial^2\zeta \dot h_{ij}^2
        +d_6\psi_{,ij}\dot h_{ik}\dot h_{jk}
        +\frac{d_7}{a^2}\zeta_{,ij}\dot h_{ik}\dot h_{jk}
        \right] \; ,
    \end{align*}
and in our case we have
\begin{equation*}
    d_4 = d_5=d_6=d_7 = 0\;.
\end{equation*}
The cubic $\zeta hh$ action also involves at most 2 spatial derivatives.

The cubic action with tensor modes only is given by
\eqref{jun11-22-30}. It  involves at most 2 spatial derivatives
as well.

\section*{Appendix D. Covariant Lagrangian functions}

In this Appendix we collect concrete expressions of covariant coupling functions $G_2$, $G_3$ and $G_4$ for the contraction stage model \eqref{A_old} with \eqref{jul18-22-1}. We have already started this procedure in Sec.~\ref{sec:dilatation}, so we start with obtained formulas \eqref{A_phi_X}. 
Substituting concrete expressions for $a_2$ and $a_3$ in eq.~\eqref{A_phi_X} we arrive to
    \begin{subequations}
    \label{A_phi_X_a_small}
    \begin{align}
      &A_2 =  \hat{g}
      \text{e}^{(2\mu+2)\phi} a_2\Big(\frac{\text{e}^{\phi}}{\sqrt{2X}}\Big) = \hat{g}
      \text{e}^{(2\mu+2)\phi} (c_2 + \sqrt{2X} d_2 \text{e}^{-\phi})\;\text{,} \\ 
        &A_3 = \hat{g} \text{e}^{(2\mu+1)\phi} a_3\Big(\frac{\text{e}^{\phi}}{\sqrt{2X}}\Big) = \hat{g}
      \text{e}^{(2\mu+1)\phi}(c_3 + \sqrt{2X} d_3 \text{e}^{-\phi})\; \text{,} \\
      &A_4 =  - \frac{\hat{g}}{2}
      \text{e}^{2\mu\phi}\; .
    \end{align}
    \end{subequations}
The relationship between the Lagrangian functions in the covariant and ADM formalisms is given
    by formulas \eqref{jan23-22-1} and \eqref{F}.
For completeness, we write an expression for $F_X$:    
    \begin{equation*}
        F_X = -\frac{\hat{g}\text{e}^{2\mu \phi}\left(\sqrt{2}c_3\text{e}^{\phi}+2(d_3+2\mu)\sqrt{X}\right)}{4X^{3/2}}\; \text{.}
    \end{equation*}
Substituting the latter formula together with \eqref{A_phi_X_a_small} into formulas \eqref{jan23-22-1}, we immediately arrive to the following covariant Lagrangian functions
\begin{subequations}
\label{app:Horndeski_func_J}
\begin{align}
    &G_2 = \hat{g}\text{e}^{2\mu \phi}\Big[\text{e}^{ \phi}\big(c_2\text{e}^{ \phi} + \sqrt{2}(d_2-c_3-2c_3\mu)\sqrt{X}\big)+2\mu(d_3+2\mu)X\text{ln}X\Big],\\
    &G_3 = \frac{1}{2}\hat{g}\text{e}^{2\mu \phi}(d_3+2\mu)(2+\text{ln}X),\\
    &G_4 = \frac{1}{2}\hat{g}\text{e}^{2\mu \phi}.
\end{align} 
\end{subequations}

Next, we find the covariant coupling functions in Einstein frame as well. To this end we will use formulas from Ref.~\cite{Rubakov:2017xzr}, Appendix A.5. Using similar notations as in Ref.~\cite{Rubakov:2017xzr}, we rewrite conformal transformation of the metric as
    \be
    \label{app:gmunu}
        g_{\mu \nu \,  (E)}\; = \text{e}^{2K(\phi)}g_{\mu\nu}, 
    \ee
We stick to the action \eqref{Hor_L} but with $G_4 \equiv G_4(\phi)$ in Jordan frame, while in Einstein frame we demand
    \begin{align}
    \label{app:horndeski_action_E}
      \cal S=&\int d^4x \sqrt{-g_{(E)}}
      \left\{ G^{(E)}_2\left(\phi, X_{(E)}\right)-G^{(E)}_3\left(\phi, X_{(E)}\right)(\Box \phi)^{(E)}
      + \frac{M_{P}^2}{2}R_{(E)} \right\}\;,
    \end{align}
where $(E)$ is just a sub- or superscript meaning Einstein frame.
One immediately writes
\begin{equation}
\label{app:XE}
    X_{(E)} = -\frac{1}{2}g_{(E)}^{\mu\nu} \partial_{\mu}\phi\partial_{\nu}\phi= \text{e}^{-2K}X,
\end{equation}
and 
\begin{equation}
    \sqrt{-g_{(E)}} = \text{e}^{4K}\sqrt{-g}.
\end{equation}
Next, one can find
\begin{equation}
\label{app:boxE}
    (\Box \phi)^{(E)} = g_{(E)}^{\mu\nu} \nabla^{(E)}_{\mu}\partial_{\nu} \phi = \text{e}^{-2K} [\Box\phi  +2g^{\mu\nu} \partial_{\mu}K\partial_{\nu}\phi],
\end{equation} 
where we use formula for the relationship between the Christoffel symbols from Ref.~\cite{Rubakov:2017xzr}.
The relation between scalar curvatures in Einstein and Jordan frame is also given in Ref.~\cite{Rubakov:2017xzr}. Finally, as a result of substituting these expressions \eqref{app:XE}-\eqref{app:boxE} in eq.~\eqref{app:horndeski_action_E} and straightforward (but
tedious) calculation we obtain 
\begin{align*}
    &G^{(E)}_2 = \frac{M_{P}^4G_2(\phi,X)}{(2G_4)^2} -  \frac{4M_{P}^4}{(2G_4)^3} G_3(\phi,X)G_{4\phi} X +12 M_{P}^2\frac{G^2_{4\phi}}{(2G_4)^3}  X , \\
    &G^{(E)}_3 = M_{P}^2\frac{G_3(\phi,X)}{2G_4(\phi)} ,\\
    &G^{(E)}_4 = \frac{M_{Pl}^2}{2}. 
\end{align*}
It is worth stressing, that during the calculation of formulas for $G^{(E)}_2$, $G^{(E)}_3$, and $G^{(E)}_4$ we obtain that function $K(\phi)$ from \eqref{app:gmunu} is
\begin{equation*}
    K(\phi)=\frac{1}{2}\text{ln}\left(\frac{2G_4}{M_{P}^2}\right),
\end{equation*}
which for sure coincides with eq.~\eqref{conf_transf}.
Finally, 
if $\frac{M_{P}^2}{2} = 1$, we arrive to
\begin{subequations}
\label{G_einst}
\begin{align}
    &G^{(E)}_2 = \frac{G_2}{G_4^2} - 2 \frac{G_{4\phi}}{G_4^3} G_3 X +3 \frac{G^2_{4\phi}}{G_4^3}  X , \\
    &G^{(E)}_3 = \frac{G_3}{G_4} ,\\
    &G^{(E)}_4 = 1.
\end{align}
\end{subequations}

Substituting Jordan frame coupling functions \eqref{app:Horndeski_func_J} into \eqref{G_einst}, we find the desired Lagrangian functions in Einstein frame
\begin{align*}
    &G^{(E)}_2 =  \frac{4\text{e}^{-2\mu \phi}}{\hat{g}} (c_2 \text{e}^{2\phi}+\sqrt{2X}d_2 \text{e}^{\phi} - \sqrt{2X}c_3(2\mu+1) \text{e}^{\phi}-4d_3\mu X - 2\mu^2 X), \\
    &G^{(E)}_3 =  (d_3 + 2\mu) (2+\text{ln}X).
\end{align*}

\section*{Appendix E. Scale factor and Hubble parameter in Einstein frame}

In this Appendix we show precise calculations of scale factor and Hubble parameter in the case of the model with $\mu=1$ from Sec.~\ref{sec:mu1} in Einstein frame. 

We remind that we proceed with the conformal transformation of the metric given by \eqref{conf_transf}. That is why, for scale factor in Einstein frame we have
\begin{equation*}
    a_E = a \sqrt{\frac{2B_4}{M_{Pl}^2}}  = (-t)^{\chi - 1} \frac{d\sqrt{g}}{M_{Pl}}  \equiv (-t)^{\chi - 1} d g_1 = \frac{g_1}{(1-\chi)(-\eta)},
\end{equation*}
where we remind that $a$ is a scale factor in Jordan frame. For the last equality we use  more accurate (than we have already written in Sec.~\ref{sec:intro}) formula for the relation between cosmic time in Jordan frame and conformal time:
\begin{equation*}
    (-t) = \big[d(1-\chi)(-\eta)\big]^{\frac{1}{1-\chi}}.
\end{equation*}
We also introduce $g_1 \equiv \frac{\sqrt{g}}{M_{Pl}}$ for simplicity. Note, that the dimension of the latter is $-1$. Next, we find cosmic time in Einstein frame
\begin{align*}
    &t_E = \int_{-\infty}^0 a_E d\eta = -\frac{g_1}{(1-\chi)} \text{ln}(-d^{1/\chi}\eta),
\end{align*}
and, vice versa
\begin{equation*}
    (-\eta) = \frac{1}{d^{1/\chi}} e^{-(1-\chi)\frac{t_E}{g_1}}.
\end{equation*}
Finally, the scale factor in Einstein frame in terms of cosmic time $t_E$ is
\begin{equation*}
    a_E  =  \frac{g_1}{(1-\chi)} d^{1/\chi} e^{(1-\chi)\frac{t_E}{g_1}}.
\end{equation*}
This scale factor $\frac{a_E(1-\chi)}{g_1d^{1/\chi}}$ as a function of $\frac{t_E(1-\chi)}{g_1}$ is shown in Fig.~\ref{fig:a_E_mu_1}.
The Hubble parameter in Einstein frame is given by
    \begin{equation*}
    H_E \equiv \frac{d a_E}{a_Ed t_E} = \frac{1-\chi}{g_1}.
\end{equation*}
As it was discussed in Sec.~\ref{sec:intro}, we indeed obtain exponential expansion in Einstein frame in this case.

\section*{Appendix F. Stable and subluminal cosmology: contraction, bounce and subsequent GR kination}

In this Appendix we give a numerical example of stable and subluminal evolution from contraction stage to bounce and then to kination. The main idea is to ``sew'' already constructed healthy bounce model from Sec. \ref{sec:mu_less_1}  with the  subsequent kination epoch from \cite{Ageeva:2021yik} (see Sec.~3.3 within),  with the contraction stage from this work given by the Lagrangian \eqref{A_old}, \eqref{a_2}, \eqref{a_3}. In this model,
the Einstein-Hilbert action for gravity is restored during the kination epoch. To this end,
following Ref.~\cite{Ageeva:2021yik}, we end up with the following example of Lagrangian functions:
\begin{subequations}
\label{A_full}
	\begin{align}
	&A_2(\hat{t},N) = U_2(\hat{t})\cdot\hat{g}\cdot f(\hat{t})^{-2\mu-2}a_2(N)+ \frac{1}{2}\big(1-U_2(\hat{t})\big) f_1(\hat{t})^{-2\mu_1 -2}\left(\frac{x(\hat{t})}{N(\hat{t})^2} + \frac{v(\hat{t})}{N(\hat{t})^4}\right)
	\;\text{,}  \\
	  &A_3 (\hat{t},N)= U_3(\hat{t})\cdot\hat{g}\cdot f(\hat{t})^{-2\mu-1}a_3(N) + \frac{1}{2}\big(1-U_3(\hat{t})\big)f_1(\hat{t})^{-2\mu_1 -1}\frac{y(\hat{t})}{N(\hat{t})^3}+\frac{V_3(\hat{t})}{N(\hat{t})^2}\; \text{,}
          \\
	  &A_4 (\hat{t})= -B_4(\hat{t}) = -\frac{1}{2}U_4(\hat{t})\cdot\hat{g}\cdot f(\hat{t})^{-2\mu}-\frac{1}{2}\big(1-U_4(\hat{t})\big)f_1(\hat{t})^{-2\mu_1}+V_4(\hat{t})\; .
	  \label{A_full_4}
	\end{align}
\end{subequations}
We construct these functions, following the same logic as in Ref.~\cite{Ageeva:2021yik}, so that  to obtain a smooth transition from contraction to bounce and then to GR kination stage.
We set from the beginning
\begin{align*}
    \hat{g} = 1\;. 
\end{align*}
Next, one should choose such $n_S$ and $\chi$ from the space of parameters in Fig.~\ref{fig:region}, which satisfy necessary conditions \eqref{cond_1} and \eqref{cond_2} from Sec.~\ref{sec:mu_less_1}. We thus set 
\begin{align*}
    n_S = 0.967\;, \quad \chi =0.4\;,
\end{align*}
and using \eqref{general_n_s} we also find
\begin{align*}
     \mu = 1.0099\;.
\end{align*}
We use the same value of $\mu_1$  as in \cite{Ageeva:2021yik}
\[
\mu_1 = 0.8 \; .
\]
The functions $U_2$, $U_3$ and $U_4$ in eqs.~\eqref{A_full} are given by
\begin{subequations}
\label{U}
\begin{align}
    &U_2(\hat{t}) = 1 + \text{ln}\left(\frac{\text{e}^{3.4s(-\hat{t}-55)}+\text{e}}{\text{e}^{3.4s(-\hat{t}-55)}+\text{e}^{2}}\right), \\
    &U_3(\hat{t}) =  1 + \text{ln}\left(\frac{\text{e}^{5s(-\hat{t}-28)}+\text{e}}{\text{e}^{5s(-\hat{t}-28)}+\text{e}^{2}}\right),\\
    &U_4(\hat{t}) =  1 + \text{ln}\left(\frac{\text{e}^{4s(-\hat{t}-50)}+\text{e}}{\text{e}^{4s(-\hat{t}-50)}+\text{e}^{2}}\right),
\end{align}
\end{subequations}
which were again constructed in the same manner as in Ref.~\cite{Ageeva:2021yik} and chosen so that to obtain stable and subluminal evolution at all times: contraction at large negative times, then bounce and kination at large positive times. In eqs.~\eqref{U} we set $s = 1/500$. Function $f(\hat{t})$ in eqs.~\eqref{A_full} is given by  
\begin{align*}
    f(\hat{t}) = \frac{1}{2}\left((-\hat{t}+500)+\frac{\text{ln}\Big(2\text{cosh}\big(s(\hat{t}-500)\big)\Big)}{s}\right)+c\;,
\end{align*}
with $c = 7$. Next, function $f_1$ is \cite{Ageeva:2021yik}
\begin{equation*}
  f_1(\hat{t})
  = \frac{c_1}{2}\left(-\hat{t}+\frac{\text{ln}\big(2\text{cosh}(s\hat{t})\big)}{s}\right) + 1 \; ,
\end{equation*}
and $c_1 =4\cdot 10^{-3}$. Functions $x(\hat{t})$, $v(\hat{t})$ and $y(\hat{t})$ are the same as in \cite{Ageeva:2021yik}
 \begin{align*}    
      x(t) &= x_0\big(1- U_x(t)\big) + \frac{4}{3((t+2000)^2+(t-5000)^2)}
      \cdot U_x(t),\\
      v(t) &= v_0\big(1 - U_x(t)\big) +
      \frac{v_2}{(|t|+2000)^5} \cdot U_x(t),\\
      y(t) &= y_0\big(1- U_y(t)\big) +
      \frac{y_2}{(|t|+2000)^5} \cdot U_y(t),
    \end{align*}
where
\begin{align*}
  U_x(t) & =
  \text{ln}\Big(\frac{\text{e}^{-1.5\cdot s\cdot(t-80)}+
    \text{e}^2}{\text{e}^{-1.5\cdot s\cdot(t-80)}+\text{e}}\Big),
  \\
  U_y(t) &= \text{ln}\Big(\frac{\text{e}^{-3.8\cdot s\cdot(t+180)}
    +\text{e}^2}{\text{e}^{-3.8\cdot s\cdot(t+180)}+\text{e}}\Big) ,
\end{align*}
with the following parameters
\begin{equation*}
        x_0 = - 1.6 \cdot 10^{-5},
        \quad
        y_0 = -1.2 \cdot 10^{-3} \; ,
\end{equation*}
\begin{equation*}
   v_0 = 5.19\cdot 10^{-6} \;  ,
  \end{equation*}
\begin{equation*}
    v_2 = 1.04\cdot 10^8, \quad y_2 = 9.6\cdot 10^{10}. 
\end{equation*}
Finally, we use two auxiliary functions $V_3$ and $V_4$ in eqs.~\eqref{A_full}
\begin{align*}
    V_3(\hat{t}) = -1.5\left[1 + \text{ln}\left(\frac{\text{e}^{5.15s(-\hat{t}-2100)}+\text{e}}{\text{e}^{5.15s(-\hat{t}-2100)}+\text{e}^{2}}\right)\right]\left[1 + \text{ln}\left(\frac{\text{e}^{5.2s(\hat{t}-70)}+\text{e}}{\text{e}^{5.2s(\hat{t}-70)}+\text{e}^{2}}\right)\right],
\end{align*}
\begin{align*}
    V_4(\hat{t}) = -1.5\cdot 10^4\left[1 + \text{ln}\left(\frac{\text{e}^{5.15s(-\hat{t}-2300)}+\text{e}}{\text{e}^{5.15s(-\hat{t}-2300)}+\text{e}^{2}}\right)\right]\left[1 + \text{ln}\left(\frac{\text{e}^{5.2s(\hat{t}-70)}+\text{e}}{\text{e}^{5.2s(\hat{t}-70)}+\text{e}^{2}}\right)\right],
\end{align*}
which are needed to proceed stable behaviour near bounce. 

Solving equations \eqref{eoms} with chosen functions \eqref{A_full} and with setting
\begin{align*}
    \kappa = 1\;, \quad c_3 = -5\;, \quad d_3 = -2\;,
\end{align*}
which is chosen so that to satisfy necessary conditions \eqref{cond_1} and \eqref{cond_2} from Sec.~\ref{sec:mu_less_1}.
One can also obtain, using formulas \eqref{kappa}, \eqref{chi_N}, and \eqref{rho} from Sec.~\ref{sec:mu_less_1}
\begin{align*}
    \rho = 0.053\;, \quad N_{\text{init}} = \frac{2}{\kappa}\left(1 + 2\mu - 2(\mu -1)\chi\right)  = 6.024\;.
\end{align*}
Having set Lagrangian functions and all parameters we  find Hubble parameter $H(\hat{t})$ and lapse function $N(\hat{t})$ as a solution. Note, that the initial conditions are 
\begin{align*}
    N(\hat{t}_1) = N_{\text{init}}, \quad H(\hat{t}_1) = \frac{\chi}{N_{init}\hat{t}_1},
\end{align*}
where $\hat{t}_1$ is some numerically suitable large negative time.
We show the behavior of Hubble parameter and lapse function at contraction, bounce, and kination in Fig.~\ref{fig:H_N}. The scalar coefficients $\mathcal{F}_S$ and $\mathcal{G}_S$ \eqref{eq:Fs_Gs_form} are shown in Fig.~\ref{fig:F_G_S} and scalar sound speed squared $u_S^2\equiv \frac{\mathcal{F}_S}{\mathcal{G}_S}$ \eqref{us} is shown in Fig.~\ref{fig:F_T} (right one); the stability and subluminality are explicit. We show tensor coefficient $\mathcal{F}_T$ \eqref{eq:Ft_Gt_form} for completeness in Fig.~\ref{fig:F_T} (left one). We recall that
$\mathcal{F}_T = \mathcal{G}_T$ \eqref{eq:Ft_Gt_form} and $c_T=1$ at all times.
\begin{figure}[h!]
\centering 
\includegraphics[width=7.5cm]{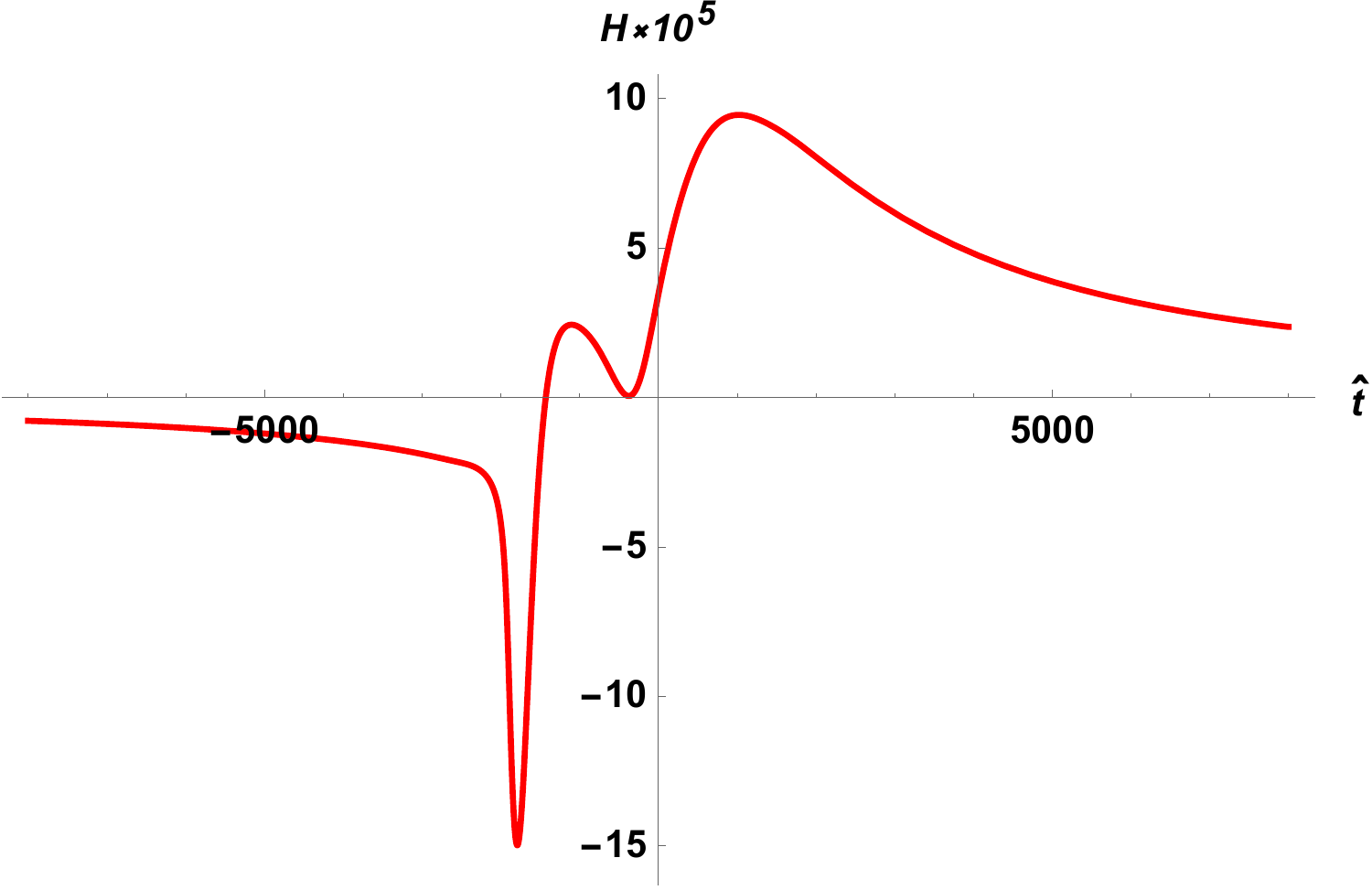}
\includegraphics[width=7.5cm]{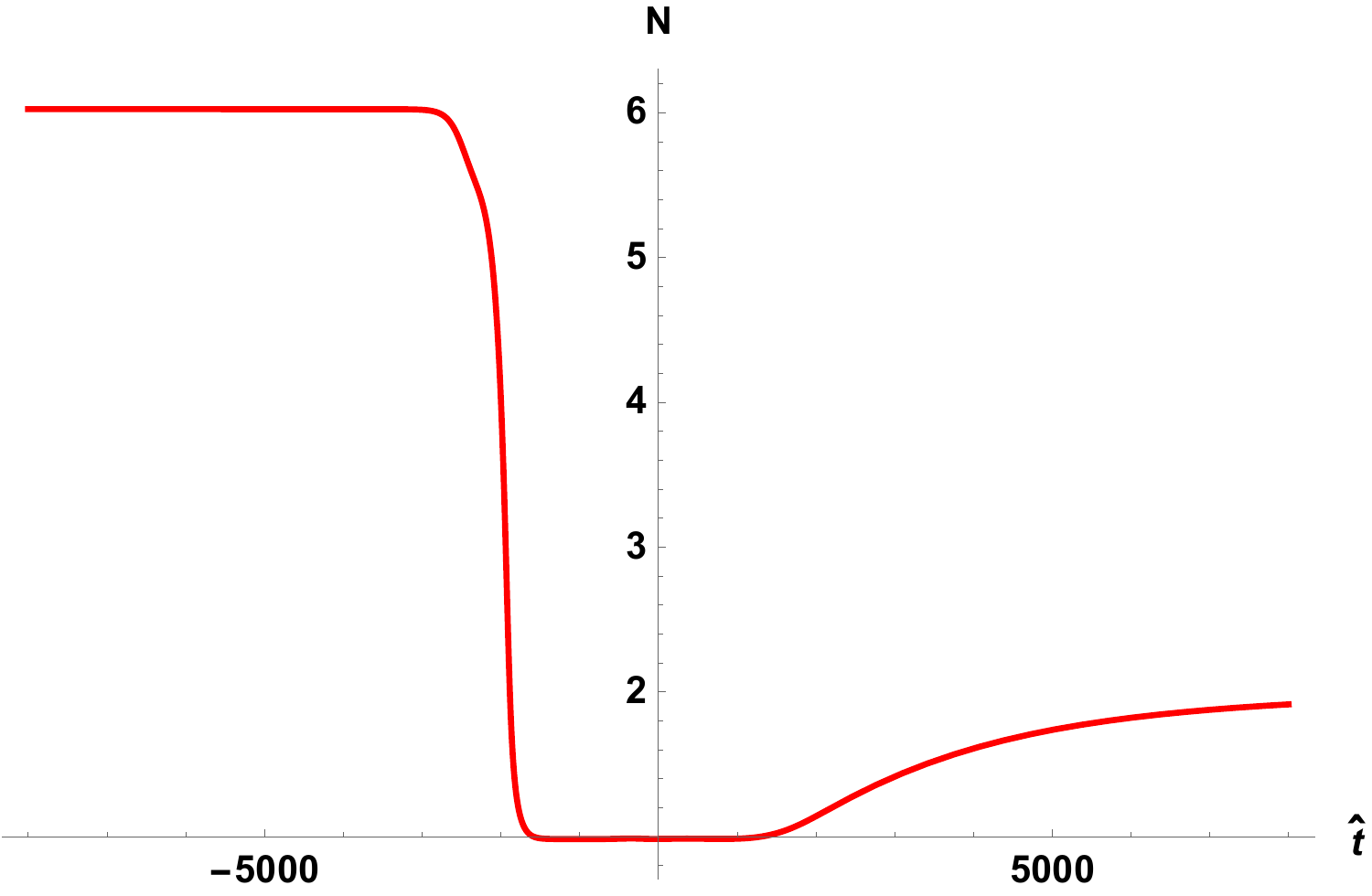}
 \caption{Hubble parameter (left panel) and lapse function (right panel) for the model of Appendix F: contraction, bounce and subsequent GR kination.}
 \label{fig:H_N}
\end{figure}

\begin{figure}[h!]
\centering 
\includegraphics[width=7.5cm]{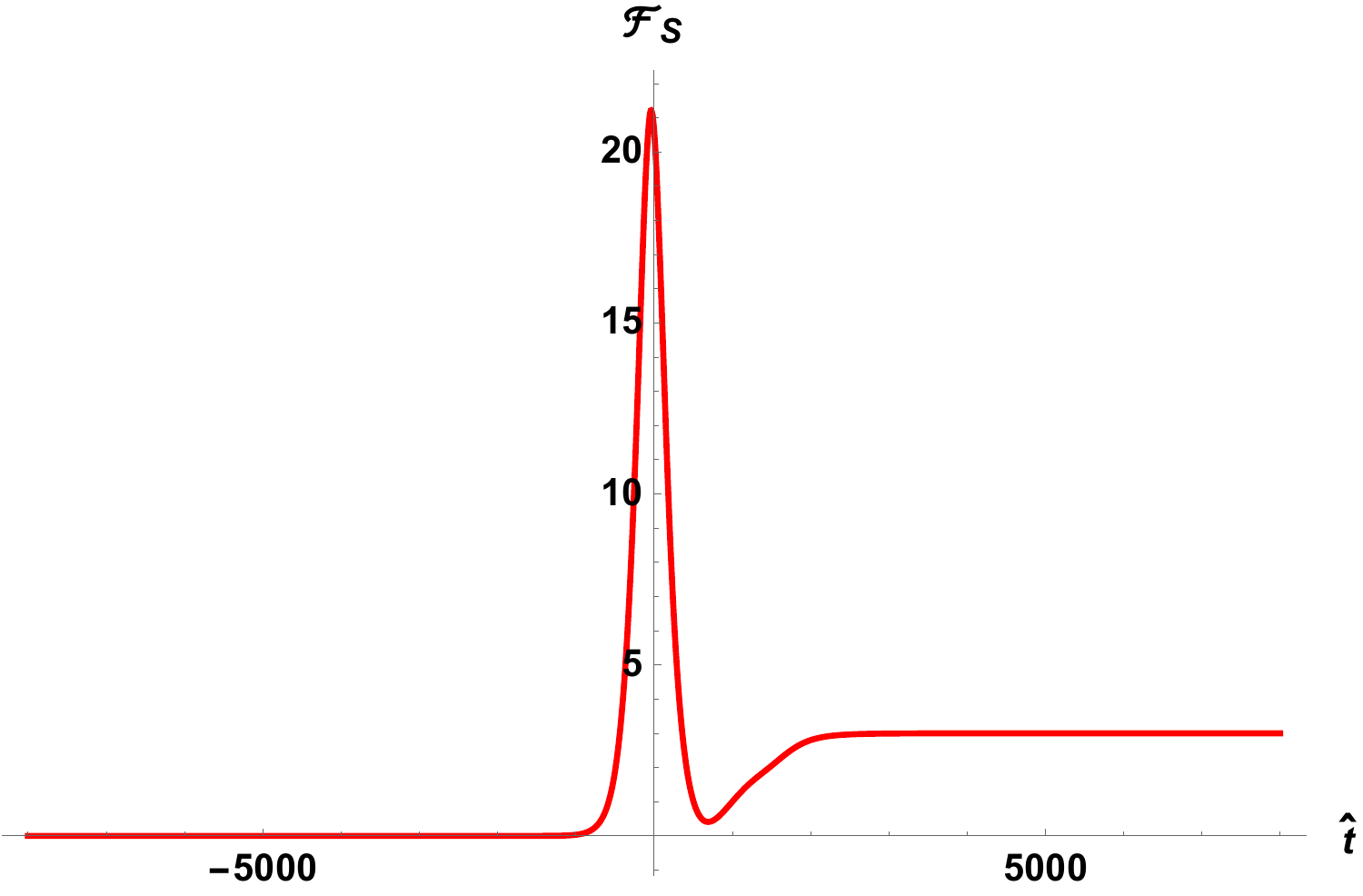}
\includegraphics[width=7.5cm]{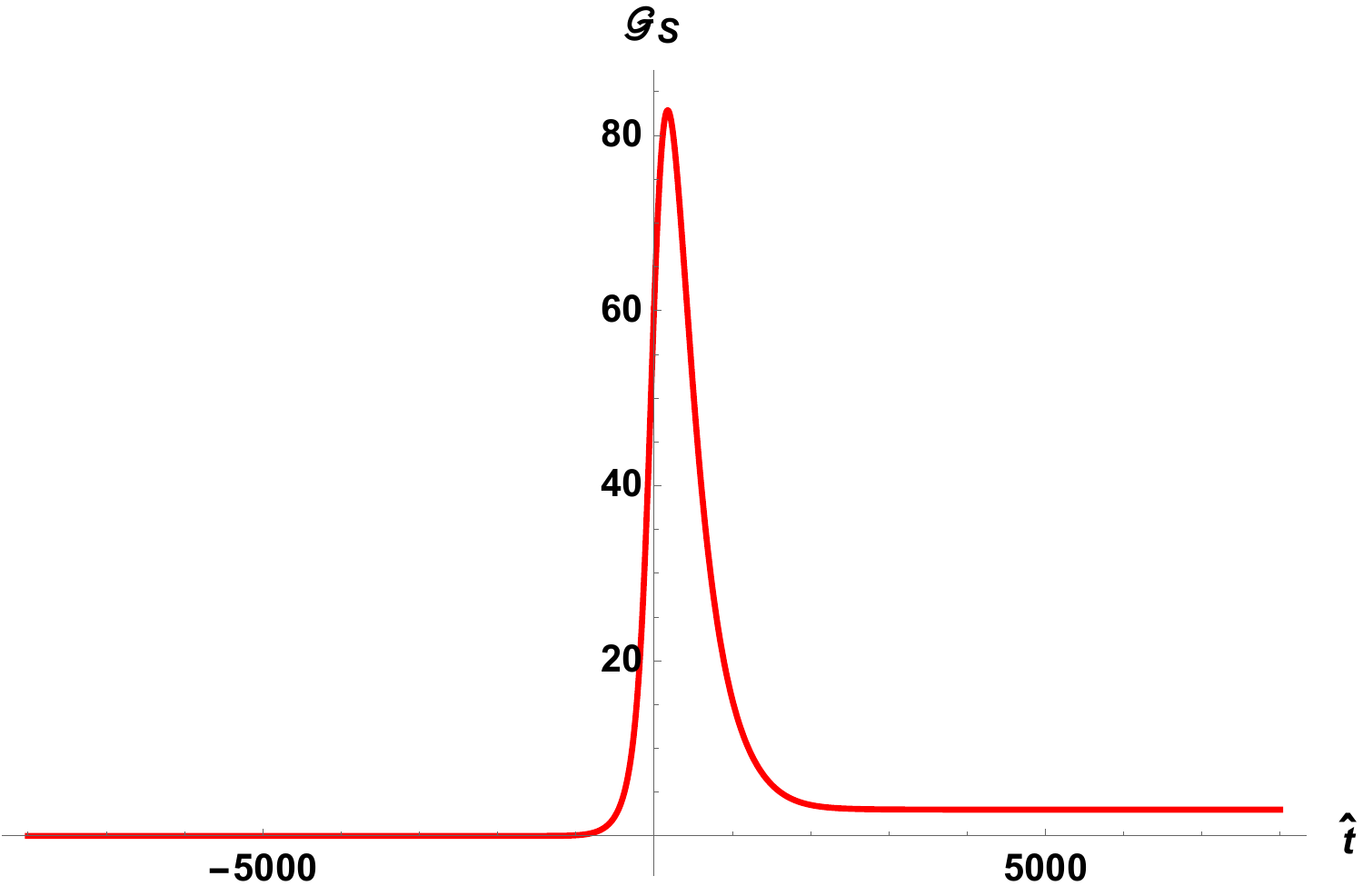}
 \caption{Coefficients $\mathcal{F}_S$ (left panel) and  $\mathcal{G}_S$ (right panel) for the model of Appendix F: contraction, bounce and subsequent GR  kination.}
 \label{fig:F_G_S}
\end{figure}

\begin{figure}[h!]
\centering 
\includegraphics[width=7.5cm]{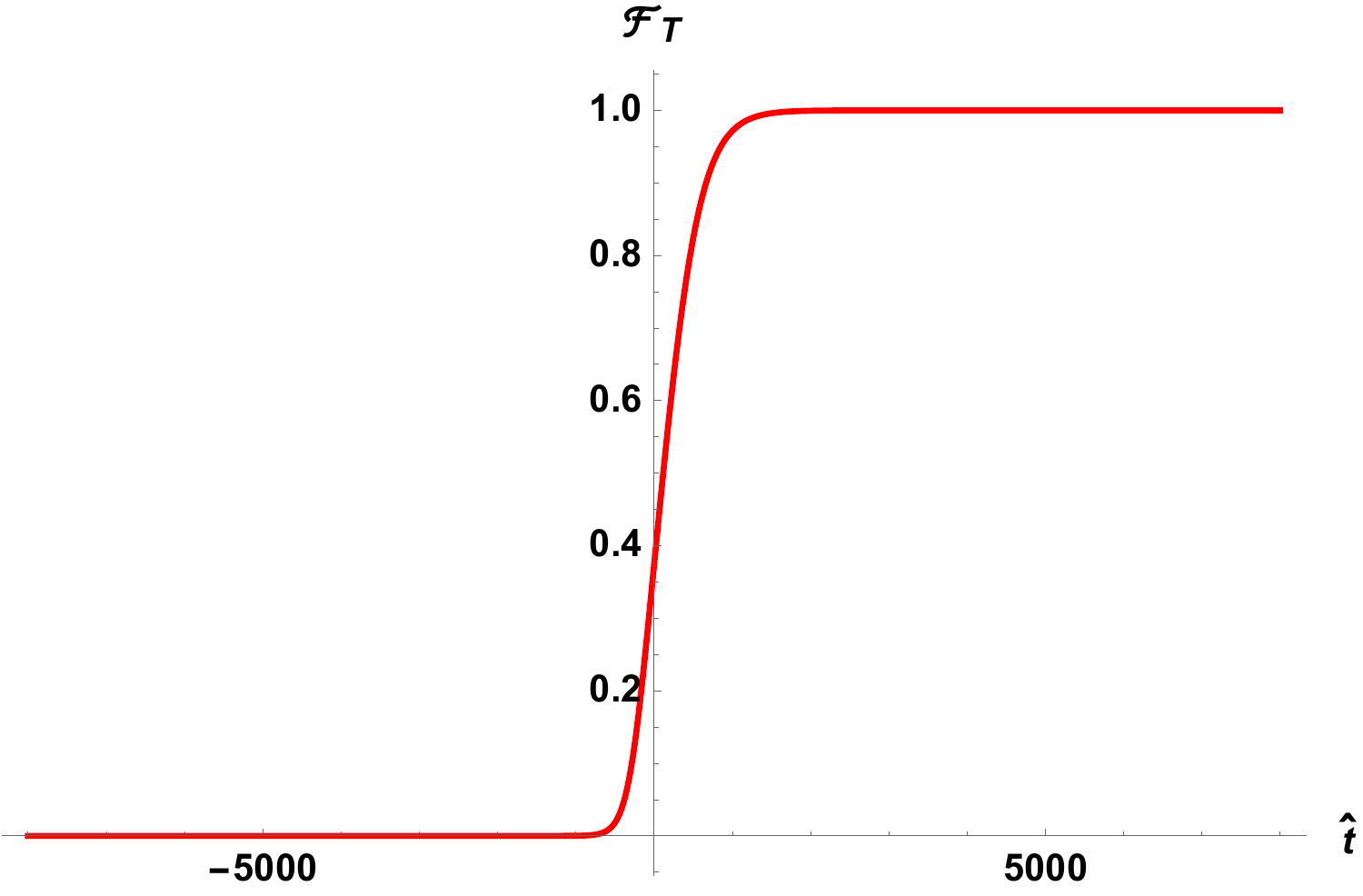}
\includegraphics[width=7.5cm]{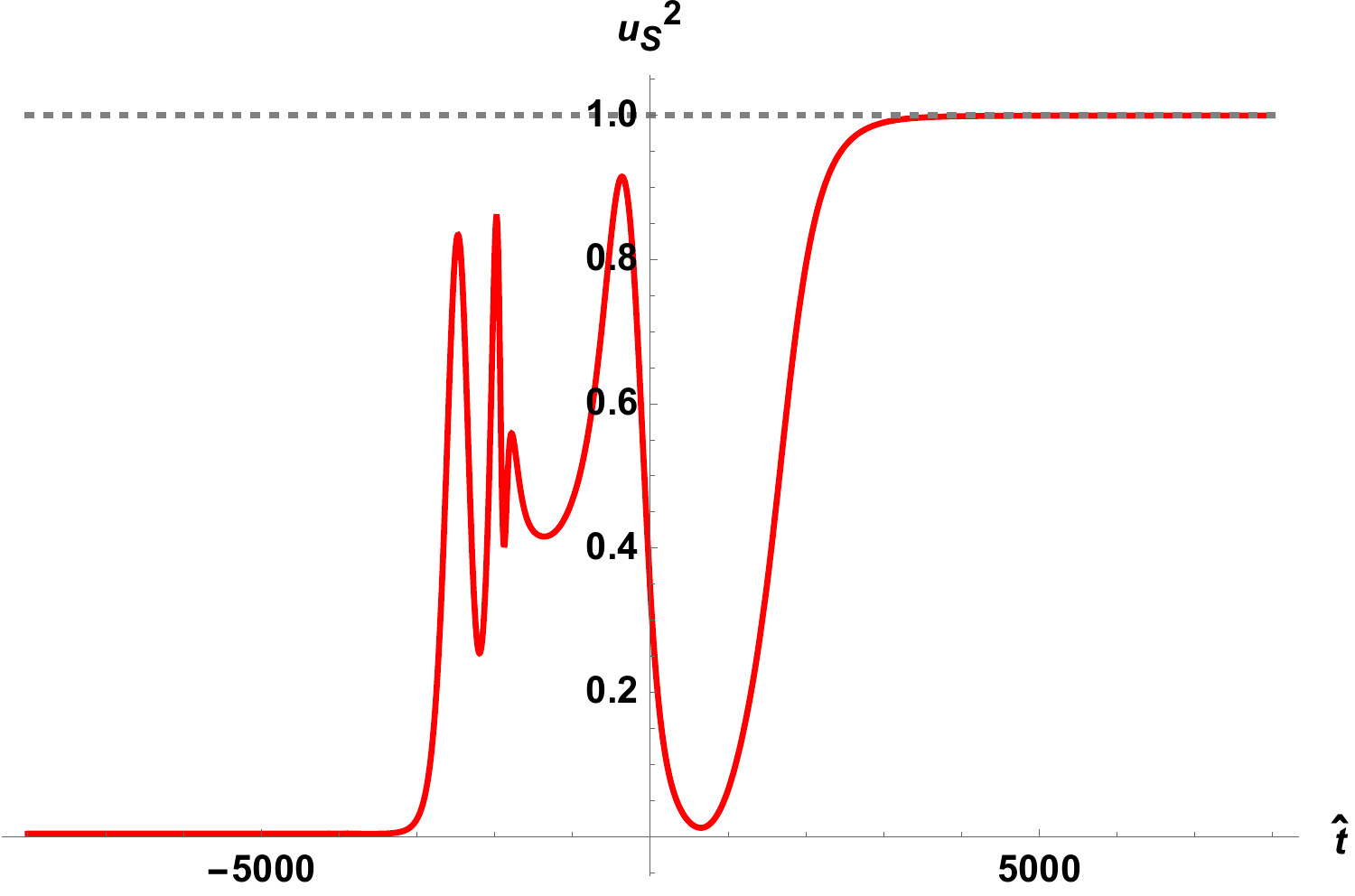}
 \caption{Coefficient $\mathcal{F}_T$ (left panel) and  $u_S^2$ (right panel) for the model of Appendix F: contraction, bounce and subsequent GR kination.}
 \label{fig:F_T}
\end{figure}

\newpage

Let us note, that at $\hat{t}\to+\infty$ we have the following asymptotic behaviour of $A_4$ in eq.~\eqref{A_full_4}
\begin{equation*}
    A_4 \to -\frac{1}{2},
\end{equation*}
i.e. we set $M_P = 1$, which means that general relativity is restored at future times in our model. 

Finally, one can make sure that the wavelength of primordial fluctuations stretches enough at the end of the contracting phase and before the universe bounces. To this end, we show when  our modes exit effective horizon
at contracting stage and  calculate the number e-folds of contraction.

Consider modes whose {\it present} wave vector is $p_0$,
roughly $p_0 \sim H_0$. Suppose, that bounce occurs at
scale factor $a=a_b$, and before that the scale factor decreased
as \eqref{jan31-22-1}.
Thus,
\be
a_b = d  |t_b|^\chi, \nonumber
\ee
where $t_b$ is the end of contraction stage. The exit of a mode
with conformal momentum $k=p_0a_0$ occurs at (see eq.~\eqref{t_f})
\be
|t_f| \sim  \left( \frac{d}{k} \right)^{\frac{1}{1-\chi}}, \nonumber
  \ee
  where we omit the dependence on sound speed $u_S$.
  Next, one can write
  \be
  k = p_0 a_0 = p_0 \frac{a_0}{a_b} d |t_b|^\chi,\nonumber
  \ee
  and obtain
  \be
  |t_f| \sim \left( \frac{1}{p_0 \frac{a_0}{a_b}  |t_b|^\chi}
    \right)^{\frac{1}{1-\chi}}.
    \nonumber
      \ee
      We now make use of the fact that at contracting
      epoch we have
      $H(t) \sim 1/|t|$ (which is valid also at $t_b$) and get
      \be
      |t_f| \sim |t_b|
      \cdot \left(\frac{H_b a_b}{p_0 a_0} \right)^{\frac{1}{1-\chi}},\nonumber
      \ee
      which, for $p_0 \sim H_0$, gives finally
      \be
       |t_f| \sim |t_b|
      \cdot \left(\frac{H_b a_b}{H_0 a_0} \right)^{\frac{1}{1-\chi}}.\nonumber
      \ee
      If $(H_0 a_0)/(H_b a_b) \ll 1$ is satisfied for the model, then horizon exit occurs long before the
      bounce. Thus, further we estimate the value of $(H_0 a_0)/(H_b a_b)$ in the model with the Lagrangian functions \eqref{A_full}. Firstly, we can immediately write the values of Hubble parameters 
      \begin{align*}
          H_0 \sim 10^2\cdot \text{km}\cdot \text{s}^{-1}\cdot\text{Mpc}^{-1} \sim 10^{-61} \;, \\
          H_b   \sim 4.2 \cdot 10^{-5} \;,
      \end{align*}
where we remind that we work in terms of $M_P =1$. Here, for $H_0$ we take some rough estimated value, and $H_b$ is obtained numerically at $\hat{t}_b \approx -2000$.\footnote{We recall that $\hat{t} = \int \frac{dt}{N(t)}$. } Note, that we define this moment of time $\hat{t}_b$ as a moment from which the formula for Hubble parameter $H\sim 1/|\hat{t}|$ becomes irrelevant. Next, we find $a_0$. It can be roughly estimated as
\begin{equation*}
    a_0 \approx a_e \frac{T_{reh}}{T_0},
\end{equation*}
where $a_e$ is the scale factor in the end of the bounce and we suppose that reheating occurs instantly after the bounce. This $a_e$ is given by
\begin{equation*}
    a_e \approx a_b \left(\int_{\hat{t}_b}^{\hat{t}_{\text{cross}}} N H d\hat{t}+\int_{\hat{t}_{\text{cross}}}^{\hat{t}_e} N H d\hat{t}\right) \equiv a_b \left(I_1+I_2\right),
\end{equation*}
where we remind that $\hat{t}_b = -2000$ and we also set $\hat{t}_e = 2000$ (approximate time when the bounce ends), while $\hat{t}_{\text{cross}}\approx-1450$ is a moment of time, where Hubble parameter changes the sign.
The integrals are calculated numerically and  equal
\begin{equation*}
    I_1 \approx -0.07, \quad I_2 \approx 0.2\;.
\end{equation*}
Finally, we write
\begin{equation*}
    \frac{H_b a_b}{H_0 a_0} \approx \frac{H_b \cdot  T_0}{H_0 \cdot  I \cdot  T_{reh}}.
\end{equation*}
Choosing, say, $T_{reh} = 10^{-12} M_P$, and having $T_0 \approx 10^{-32} M_{P}$ we arrive to
\begin{equation*}
    \frac{H_b a_b}{H_0 a_0} \approx  10^{37} \gg 1,
\end{equation*}
so, we have proved that the horizon exit occurs long before the bounce. 

Let us estimate the number of e-folds $\mathcal{N}$ which can be gained during contraction epoch. This is given by
\begin{equation*}
    \text{e}^{\mathcal{N}} \equiv \frac{a(t_f)}{a_b} = \frac{|t_f|^{\chi}}{|t_b|^{\chi}} = \left(\frac{H_b a_b}{H_0 a_0} \right)^{\frac{\chi}{1-\chi}},
\end{equation*}
so
\begin{equation*}
    \mathcal{N}  = \frac{\chi}{1-\chi}\text{ln}\left(\frac{H_b a_b}{H_0 a_0} \right) \approx 58\;,
\end{equation*}
so our model is physically relevant, i.e. the wavelength of primordial fluctuations stretches $\sim$60  Hubble scales at the end of the contracting phase and before the universe bounces.

\end{document}